\documentclass[11pt]{article}

\usepackage{jheppub} 
\usepackage[normalem]{ulem}
\usepackage[T1]{fontenc}
\usepackage{slashed,amsmath,amsfonts,amssymb,bbm}
\usepackage{mathrsfs,wasysym}
\usepackage{units}
\usepackage{graphicx}
\usepackage{makecell}
\usepackage{booktabs,multirow}
\usepackage{commath}
\usepackage{xspace}
\usepackage[dvipsnames]{xcolor} 
\usepackage{comment}
\usepackage{bbm}
\definecolor{darkblue}{rgb}{0,0,0.5}
\definecolor{darkgreen}{rgb}{0.0,0.5,0.2}
\definecolor{darkred}{rgb}{0.6,0,0}
\usepackage{hyperref}
\usepackage[capitalise]{cleveref}
\usepackage{orcidlink}

\renewcommand{\to}{\rightarrow}

\newcommand{\Ux}{$U(1)_X$\xspace}
\newcommand{\dmmodel}{DAHM-DM\xspace}
\newcommand{\cl}{C.L.\xspace}

\usepackage{tikz}
\usepackage{tikz-feynman}

\tikzfeynmanset{
yourboson/.style={
decoration={coil,aspect=0,amplitude=1.5mm,segment length=4mm,pre,pre length=0pt, post,post length=0pt},decorate}
}

\tikzfeynmanset{
yourboson2/.style={
decoration={coil,aspect=0,amplitude=1.5mm,segment length=4.1mm,pre,pre length=0pt, post,post length=0pt},decorate}
}

\subheader{\rm IFT-UAM/CSIC-25-78}

\title
{
GeV-scale thermal dark matter from dark photons: tightly constrained, yet allowed
}

\author[a,b]{D.~Alonso-González\,\orcidlink{0000-0002-7572-9184},}
\author[b]{D.~Cerde\~no\,\orcidlink{0000-0002-7649-1956},}
\author[b]{P.~Foldenauer\,\orcidlink{0000-0003-4334-4228},}
\author[a,b]{J.~M. No\,\orcidlink{0000-0002-1321-8960}}
\affiliation[a]{Departamento de F\'isica Te\'orica, Universidad Aut\'onoma de Madrid, Cantoblanco,\\ E-28049, Madrid, Spain}
\affiliation[b]{Instituto de F\'isica Te\'orica IFT-UAM/CSIC, 
Cantoblanco, E-28049, Madrid, Spain}
\emailAdd{david.alonsogonzalez@uam.es}
\emailAdd{davidg.cerdeno@gmail.com}
\emailAdd{patrick.foldenauer@csic.es}
\emailAdd{josemiguel.no@uam.es}

\abstract
{
GeV-scale thermal dark matter (DM) is highly constrained by the null results of both direct and indirect detection experiments, especially in the context of simplified models. 
In this work, we study the interplay of collider, direct and indirect detection constraints on an extension of the dark Abelian Higgs model that includes a Dirac fermionic DM candidate, $\chi$. We take into account in a consistent fashion the dilution of the indirect and direct detection signals when the relic abundance of $\chi$ is smaller than the total observed DM density (assuming that it is a subdominant component in those cases). As a consequence, we show that indirect detection constraints cannot probe regions with large kinetic mixing, and direct detection experiments provide the leading constraints in most of the parameter space.  Collider searches for the (invisibly decaying) vector mediator provide complementary bounds in areas with large kinetic mixing. We find that the only way to avoid both indirect and direct detection limits is in narrow windows of parameter space close to  $m_\chi\lesssim m_{Z_D}/2$, when $\chi$ is produced resonantly in the early universe, and it can constitute all of the DM. For this to happen, a small dark sector coupling is required: $\alpha_D~\lesssim10^{-3}$ for DM masses below $6$~GeV, or $\alpha_D~\lesssim10^{-5}$ for DM masses larger than $10$~GeV. The remaining areas of the parameter space can be probed in a complementary way by future direct detection experiments (which will narrow down the allowed area around the resonant region) and collider searches (which will set limits for smaller values of the kinetic mixing).
}

\begin{document}

\notoc

\maketitle

\section{Introduction}
\label{sec:intro}

Understanding the particle nature of dark matter (DM) remains one of the most pressing challenges in modern physics, with direct implications for physics beyond the Standard Model (SM). In recent years, models involving a hidden (or {\em dark}) sector have gained considerable attention, mostly due to their interesting phenomenology and the potential observation in upcoming experiments. In these scenarios the DM particle resides in a secluded sector, only connected to the SM through new {\em mediator} fields. Such is the case of dark photon models, where the mediator is a new gauge boson, associated to a new dark $U(1)_X$ symmetry under which the SM fields are uncharged.

In this work, we consider a minimally consistent Dark Abelian Higgs Model (DAHM) as a portal to DM. We will refer to this scenario as \dmmodel. While new vector mediators, $Z_D$, coupled to DM have been extensively studied in the literature, these works typically focus on either the light mediator regime, $m_{Z_D}\lesssim 1$~GeV \cite{Pospelov:2007mp,Pospelov:2008jk,Arkani-Hamed:2008hhe,Essig:2011nj,Frandsen:2011cg,Knapen:2017xzo,Feng:2017drg,Dutra:2018gmv,Jung:2020ukk,Fitzpatrick:2020vba,Rizzo:2020jsm,Gori:2022vri,Antel:2023hkf,Rizzo:2024bhn,Balan:2024cmq,Alenezi:2025kwl, Krnjaic:2025noj}, or the heavy mediator regime with $m_{Z_D}\gg m_{Z}$ (where high-energy collider searches are sensitive)~\cite{Feldman:2007wj,Cassel:2009pu,Dudas:2009uq,Chun:2010ve,Arcadi:2013qia,Ko:2014uka,Kahlhoefer:2015bea,Duerr:2016tmh}.
However, there is a strategic gap in the literature for dedicated studies of invisibly decaying dark photons with masses in the range of $\sim10$~GeV $\lesssim m_{Z_{D}}\lesssim m_{Z}$, above the sensitivity of $B$-factories and below the typical sensitivity of LHC searches and electroweak precision observables. Our aim is to chart this sensitivity gap by exploiting the complementarity of collider, direct detection (DD), and indirect detection (ID) constraints on a Dirac fermion, $\chi$, charged under the dark \Ux symmetry.

In recent years, DM direct detection experiments have made substantial progress in increasing their volume and lowering their backgrounds, resulting in significantly stronger upper bounds on the DM-nucleus scattering cross-section of GeV-mass candidates~\cite{XENON:2020gfr,DarkSide-50:2022qzh,PandaX-4T:2021bab,PandaX:2022aac,PandaX:2024qfu, LZ:2022lsv,LZ:2024zvo,SuperCDMS:2022zmd,SuperCDMS:2020aus,DAMIC:2020cut,CRESST:2024cpr,DEAP:2019yzn,NEWS-G:2017pxg}. These searches have already ruled out large regions of parameter space for GeV to TeV DM masses, especially for thermally produced candidates. Likewise, indirect searches for late-time DM annihilation have put pressure on the thermal freeze-out scenario. In particular, analyses of the observed temperature and polarization power spectra of the cosmic microwave background (CMB) as well as the gamma ray emission from dwarf spheroidal galaxies place strong bounds on GeV-scale thermal candidates~\cite{Galli:2011rz,Gaskins:2016cha,Leane:2018kjk,Hess:2021cdp,Kawasaki:2021etm,Dutta:2022wdi,Gajovic:2023bsu,ODonnell:2024aaw,Wang:2025tdx}. These are also complementary to accelerator searches, which either aim to produce and observe the DM particle itself, or the mediator between the DM and the SM visible sector. In the context of invisibly decaying dark photons, the leading collider probes are searches for events in which a visible object (e.g., a photon or a jet) is recoiling against a sizeable amount of missing transverse momentum, $p_{T}^{\rm miss}$, (commonly referred to as mono-$X$ searches)~\cite{CMS:2020krr,ATLAS:2021shl,ATLAS:2024rcx,ATLAS:2024rlu,ATLAS:2024xne,CMS:2025izi,CMS:2025idq}, as well as precision measurements of electroweak (EW) observables.

Combining current leading direct and indirect DM searches, GeV-scale thermal relics appear to be excluded up to masses of several tens of GeV~\cite{Billard:2021uyg,Cooley:2022ufh}. However, care has to be taken when applying these bounds to a specific model, since the relic density is a function of the input parameters. This can result in regions of the parameter space where the cosmological abundance of a particular DM candidate, $\Omega h^2_{\chi}$, is smaller than the total observed cold DM relic abundance, $\Omega h^2_{\rm CDM}$ (in such a case, DM would be multicomponent). If we denote by $\xi=\Omega h^2_{\chi}/\Omega h^2_{\rm CDM}$ the amount by which the DM candidate $\chi$ is under-abundant, for consistency, we must apply the same ratio to the local DM density and to the galactic density profiles. This dilution factor affects the observables of direct and indirect detection differently. On the one hand, for direct detection, the detection rate is proportional to the local DM density (of that particular DM candidate), which is suppressed by $\xi$. On the other hand, indirect detection processes related to DM pair annihilation, would be suppressed by $\xi^2$. This can result in a relaxation of existing limits, especially for indirect detection observables. This, of course, does not apply to collider searches.

In this work, we study the interplay of collider, direct and indirect detection constraints on an extension of the DAHM by fermionic DM candidate $\chi$, taking into account in a consistent fashion the impact of the dilution factor $\xi$ on these searches. This article is organised as follows. In~\cref{sec:model}, we introduce the theoretical framework of the Hidden Abelian vector portal model with a fermionic DM candidate. In~\cref{sec:constraints}, we give a detailed analysis of the relevant constraints from direct and indirect detection, as well as collider searches. Finally, we discuss our results in~\cref{sec:res} before presenting our conclusions in \cref{sec:conclusions}.

\section{Hidden Abelian vector
portal to dark matter}
\label{sec:model}

In this section, we will review in detail the theoretical setup of the dark Abelian Higgs model extended by a fermionic DM candidate, or \dmmodel for short.

\subsection{Minimal Dark Photon extension of the SM}
\label{sec:2.1}

We begin by considering a consistent setup for minimal $U(1)$ kinetic mixing.
Our starting point is the general dark photon Lagrangian in the gauge basis, given by (see, e.g., Ref.~\cite{Bauer:2022nwt}) 
\begin{align}
    \mathcal{L} =& - \frac{1}{4}\,
    (\hat B_{\mu\nu}, \hat W^3_{\mu\nu}, \hat X_{\mu\nu})
    \begin{pmatrix}
    1 & 0 & \epsilon \\
    0 & 1 & 0 \\
    \epsilon & 0 & 1
    \end{pmatrix}
    \begin{pmatrix}
    \hat B^{\mu\nu}\\
    \hat W^{3\mu\nu}\\
    \hat X^{\mu\nu}
    \end{pmatrix} 
    - \left( g'\, j^Y_\mu , g\, j^3_\mu, \hat{g}_D \,j_\mu^X \right) 
    \begin{pmatrix}
    \hat B^{\mu}\\
    \hat W^{3\mu} \\ 
    \hat X^{\mu}
    \end{pmatrix} 
    \nonumber \\
    &+  \frac{1}{2} \, 
    (\hat B_{\mu}, \hat W^{3}_\mu, \hat X_{\mu})\, \frac{v^2}{4}
    \begin{pmatrix}
    g'^2 & - g' g & 0 \\
   - g' g & g^2 & 0 \\
    0 &  0 & \frac{4\,\hat{M}_{X_0}^2}{v^2}
    \end{pmatrix}
    \begin{pmatrix}
    \hat B^{\mu}\\
    \hat W^{3\mu} \\ 
    \hat X^{\mu}
    \end{pmatrix}  \, , 
    \label{eq:loop_lag}
\end{align}
where the hatted fields $\hat B^{\mu}$ and $\hat W^{3\mu}$ denote the hypercharge $U(1)_Y$ and neutral $SU(2)_L$ gauge bosons of the SM, $\hat X^{\mu}$ is the gauge boson of the dark \Ux symmetry, and $\epsilon$ corresponds to the kinetic mixing between \Ux and $U(1)_Y$ (leading to non-diagonal kinetic terms). The coupling constants of $U(1)_Y$ and $SU(2)_L$ are $g'$ and $g$, respectively, and $j^Y_\mu, j^3_\mu$ their corresponding gauge (fermionic) currents. Similarly, $\hat g_D$ denotes the coupling constant and $j_\mu^X$ the dark fermionic current of \Ux (see \cref{sec:darkfermion} for more details). Finally, $v$ denotes the SM Higgs vacuum expectation value (VEV) and $\hat{M}^2_{X_0}$ is the bare tree-level mass term of the \Ux gauge boson, whose origin we discuss in \cref{sec:2.2}.

We first diagonalize the upper block-diagonal mass matrix of~\cref{eq:loop_lag} via the standard weak rotation,\footnote{Alternatively, one can start by diagonalizing the kinetic terms in~\cref{eq:loop_lag} via the transformation $G = \begin{pmatrix}
 1 & 0 & -\epsilon/\sqrt{1-\epsilon^2} \\
 0 & 1 & 0 \\
 0 & 0 & 1/\sqrt{1-\epsilon^2} 
\end{pmatrix}$, and then fully diagonalize the resulting mass matrix for the gauge bosons (now with diagonal kinetic terms).}
\begin{equation}
  R_W = 
\begin{pmatrix}
 c_W & - s_W & 0 \\
s_W & c_W & 0 \\
 0 & 0 & 1
\end{pmatrix}\, ,
\end{equation}
where $s_W$ ($c_W$) is the sine (cosine) of the weak mixing angle. We conveniently redefine the fermionic currents in~\cref{eq:loop_lag} after the weak rotation to read
\begin{equation}
    \mathcal{L} \supset - \left(  e\, j^\mathrm{EM}_\mu , g_Z\, j^Z_\mu, \hat{g}_D \,j_\mu^X \right) 
    \begin{pmatrix}
    \hat A^{\mu}\\
    \hat Z^{\mu} \\ 
    \hat X^{\mu}
    \end{pmatrix}\, ,
\end{equation}
where $\hat A^{\mu}$ and $\hat Z^{\mu}$ are the would-be physical photon and $Z$-boson in the absence of kinetic mixing, and $j^\mathrm{EM}_\mu$, $ j^Z_\mu$ their respective associated fermionic currents in the SM.
The resulting kinetic terms for $\hat A^{\mu}$, $\hat Z^{\mu}$ and $\hat X^{\mu}$ are diagonalized through the non-unitary transformation
\begin{equation}
  \begin{pmatrix}
    \hat A^{\mu}\\
    \hat Z^{\mu} \\ 
    \hat X^{\mu}
    \end{pmatrix} = 
\begin{pmatrix}
 1 & 0 & \frac{- \epsilon\, c_W}{ \sqrt{1-\epsilon^2}} \\
 0 & 1 & \frac{\epsilon\, s_W}{ \sqrt{1-\epsilon^2}} \\
 0 & 0 & \frac{1}{ \sqrt{1-\epsilon^2}} 
\end{pmatrix}\, 
\begin{pmatrix}
    A^{\mu}\\
    \tilde Z^{\mu} \\ 
    \tilde X^{\mu}
    \end{pmatrix}\, ,
\end{equation}
upon which the fermionic currents and gauge mass terms in~\cref{eq:loop_lag} become 
\begin{align}
\mathcal{L} \supset& \frac{1}{2} \, 
    (A_{\mu}, \tilde Z_\mu, \tilde X_{\mu})\,
    \begin{pmatrix}
    0 & 0 & 0 \\
    0 & M_{Z_0}^2 & M_{Z_0}^2 \frac{s_W \epsilon}{\sqrt{1-\epsilon^2}} \\
    0 &  M_{Z_0}^2\frac{s_W \epsilon}{\sqrt{1-\epsilon^2}} & \frac{\hat{M}_{X_0}^2 +  M_{Z_0}^2 s_W^2 \epsilon^2}{1-\epsilon^2}
    \end{pmatrix}
    \begin{pmatrix}
    A^{\mu}\\
    \tilde Z^\mu \\
    \tilde X^{\mu}
    \end{pmatrix} \nonumber  \\
    &- \left( e\, j^\mathrm{EM}_\mu , g_Z\, j^Z_\mu , \hat{g}_D \,j_\mu^X \right)
    \begin{pmatrix}
    A^{\mu} -  \frac{c_W \epsilon \tilde X^{\mu}}{\sqrt{1-\epsilon^2}} \\
    \tilde Z^\mu + \frac{s_W \epsilon \tilde X^{\mu}}{\sqrt{1-\epsilon^2}} \\
    \frac{\tilde X^{\mu}}{\sqrt{1-\epsilon^2}}
    \end{pmatrix} \, , 
    \label{eq:loop_lag_diag}
\end{align}
with $M_{Z_0}^2 = (g^2 + g'^2)\, v^2/4$. From~\cref{eq:loop_lag_diag}, it is apparent that $A_\mu$ is already the true, massless photon. We can also redefine the dark gauge coupling to $g_D = \hat g_D/\sqrt{1-\epsilon^2}$, which slightly simplifies~\cref{eq:loop_lag_diag} to 
\begin{align}
\mathcal{L} \supset& \frac{1}{2} \, 
    (A_{\mu}, \tilde Z_\mu, \tilde X_{\mu})\,
    \begin{pmatrix}
    0 & 0 & 0 \\
    0 & M_{Z_0}^2 & M_{Z_0}^2 \delta \\
    0 &  M_{Z_0}^2 \delta & M_{X_0}^2 +  M_{Z_0}^2 \delta^2
    \end{pmatrix}
    \begin{pmatrix}
    A^{\mu}\\
    \tilde Z^\mu \\
    \tilde X^{\mu}
    \end{pmatrix}
    \nonumber  \\
    &- \left( e\, j^\mathrm{EM}_\mu , g_Z\, j^Z_\mu , g_D \,j_\mu^X \right)
    \begin{pmatrix}
    A^{\mu} -  \frac{c_W \delta \tilde X^{\mu}}{s_W} \\
    \tilde Z^\mu + \delta \tilde X^{\mu} \\
    \tilde X^{\mu}
    \end{pmatrix} \, ,
        \label{eq:loop_lag_diag2}
\end{align}
with $\delta^2 = s_W^2\epsilon^2/(1-\epsilon^2)$ and $M_{X_0}^2 = \hat{M}_{X_0}^2/(1-\epsilon^2)$. The mass matrix in~\cref{eq:loop_lag_diag2} may be fully diagonalized by a rotation of angle $\alpha$ to the physical $Z$ boson and dark photon $Z_D$ (with masses $m_Z$ and $m_{Z_D}$) via
\begin{equation}\label{eq:phys_diag}
    \begin{pmatrix}
    \tilde Z^\mu \\
    \tilde X^{\mu}
    \end{pmatrix} = \begin{pmatrix}
    c_\alpha & s_\alpha \\
    -s_\alpha & c_\alpha
    \end{pmatrix} \begin{pmatrix}
    Z^\mu \\
    Z_D^{\mu}
    \end{pmatrix} \,,
\end{equation}
with $c_\alpha \equiv {\rm cos}\, \alpha$, $s_\alpha \equiv {\rm sin}\, \alpha$. In the rest of this work, we assume $m_{Z_D} < m_{Z}$ (i.e., we focus on light dark photons). After a bit of algebra, the mixing $s_\alpha$ can be written in terms of the kinetic mixing parameter $\epsilon$ and the ratio of physical masses $R \equiv m_{Z_D}^2/m_{Z}^2 < 1$ as 
\begin{equation}
    s_\alpha^2 =  \frac{1}{2 (1+\delta^2)} \times \left[\frac{2\,\delta^2 }{1-R} + 1 - \sqrt{ 1 - \frac{4 \,R \, \delta^2 }{(1-R)^2} } \right]\, .
    \label{gauge_mixing_angle}
\end{equation}
Consistency of~\cref{gauge_mixing_angle} 
requires $\delta^2 \leq (1-R)^2/(4R)$. This is simply a consequence of level-splitting, due to which the mass eigenvalues cannot be arbitrarily close to each other for a given $\epsilon \neq 0$. 
For small kinetic mixing $\epsilon \ll 1$,~\cref{gauge_mixing_angle} may be Taylor-expanded, yielding a simple approximate relation between the gauge mixing $s_{\alpha}$ 
and the kinetic mixing parameter $\epsilon$
\begin{align}
     s_\alpha  & =  -\eta \,\epsilon + \mathcal{O}(\epsilon^3)\, ,\nonumber \\ 
    c_\alpha  &= 1 - \frac{\eta^2}{2}\,\epsilon^2 + \mathcal{O}(\epsilon^4) \, ,
\label{eq:gauge_mixing_angle_Taylor}
\end{align}
with $\eta = s_W/(1-R)$. This highlights that the mixing $s_\alpha$ of $Z$ and $Z_D$ is of the same order of magnitude as (and directly proportional to) the kinetic mixing $\epsilon$, except when $m_{Z_D} \to  m_{Z}$, in which case the mixing $s_\alpha$ can be much larger than $\epsilon$.

\subsection{Dark Abelian Higgs extension of the SM}
\label{sec:2.2}

Turning now to the mass-generation mechanism for the dark gauge boson, $\hat X_{\mu}$, we consider that it becomes massive by the spontaneous breaking of the $U(1)_X$ symmetry, driven by the VEV of a complex scalar field $S$ (a dark-Higgs field, singlet under the SM and charged under the $U(1)_X$ symmetry). This dark Abelian Higgs model has been thoroughly studied in the literature (see, e.g., Refs.~\cite{Schabinger:2005ei,Gopalakrishna:2008dv,Wells:2008xg,Curtin:2014cca}). The singlet scalar couples to the SM via a Higgs portal term $\left|H \right|^2 S^* S$. When both the singlet $S$ and the Higgs field $H$ develop their corresponding VEVs, this portal term leads to mixing between the SM Higgs boson and the dark Higgs boson. This establishes an additional link between the visible and dark sectors, in addition to the gauge kinetic mixing discussed in the previous section. 
The general scalar potential reads
\begin{equation}
V(H,S) =  - \mu^2 \, H^\dagger H + \lambda (H^\dagger H)^2  -\mu_s^2\, S^* S +{\lambda_s} (S^* S)^2  + \lambda_{m}\, S^* S \, H^\dagger H\,.
\end{equation}
After EW and $U(1)_X$ symmetry breaking, we can express the scalar fields in unitary gauge as
\begin{align}
    H &= \frac{1}{\sqrt{2}} (v + h)\,,\nonumber\\ 
    S &= \frac{1}{\sqrt{2}} (w + s)\,,
\end{align}
where $v = 246$~GeV and $w$ are the respective EW and singlet VEVs, $h$ is the SM Higgs boson, and $s$ is the $U(1)_X$ Higgs boson. The scalar mass matrix for $(h,\, s)$ reads
 \begin{equation}
    \mathcal{M}^2 = \left(\begin{matrix}
    m_h^2 & m_{hs}^2 \\
    m_{hs}^2 &   m_s^2
    \end{matrix}\right) =
    \left(\begin{matrix}
    2 \lambda v^2 & \lambda_{m} v w \\
    \lambda_{m} v w &   2 \lambda_s w^2
    \end{matrix}\right) \,,
\end{equation}
with the term $m_{hs}^2 \propto \lambda_{m}$ inducing mixing between $h$ and $s$. 
The squared-mass matrix $\mathcal{M}^2$ can be diagonalized by a simple rotation of angle $\theta$, leading to the mass eigenstates $h_{125},\,h_D$:
 \begin{equation}
 \label{Higgs_eigenstates}
 \left(\begin{matrix}
h_{125}\\
h_D
  \end{matrix}\right) 
  =
   \left(\begin{matrix}
c_\theta & \, - s_\theta \\
s_\theta &  c_\theta
  \end{matrix}\right) \,  \left(\begin{matrix}
h\\
s
  \end{matrix}\right) \,,
\end{equation}
with the abbreviations $s_\theta = \sin \theta$ and $c_\theta = \cos \theta$, and $h_{125}$ being the 125 GeV Higgs boson and $h_D$ the dark (singlet-like) Higgs boson. The angle $\theta$ is given by (see, e.g., Ref.~\cite{Wells:2008xg})
\begin{equation}
\tan {2 \theta} = \frac{\lambda_{m} v w}{\lambda_s w^2 - \lambda v^2}\,.
\end{equation}

The mass parameter $\hat{M}_{X_0}$ for the $U(1)_X$ gauge boson $\hat{X}_\mu$ (recall \cref{eq:loop_lag}) is given in terms of the singlet VEV $w$, the dark gauge coupling $\hat{g}_D$ and $U(1)_X$ charge $q_S$ of the Abelian Higgs field $S$ (which we set to $q_S = 1$ for the rest of the article) as
\begin{equation}
\hat{M}_{X_0} = \frac{q_S \, \hat{g}_D\, w}{2} \quad \longrightarrow  \quad M_{X_0} = \frac{q_S \, g_D\, w}{2}\, .
\end{equation}

Considering now the interactions between the scalars and physical gauge vector-bosons $Z$, $Z_D$, they can be obtained from \cref{eq:loop_lag_diag2} by replacing $M_{Z_0}^2 \to M_{Z_0}^2 (1 + h/v)^2$ and $M_{X_0}^2 \to M_{X_0}^2 (1 + s/w)^2$. After some algebra, and using also~\cref{Higgs_eigenstates}, we get for the scalar-vector-vector interactions:
\begin{align}
\label{eq:DAHM_interactions}
h_{125}\,Z^\mu Z_{\mu} &\longrightarrow \frac{2 m_Z^2}{v} \left[c_\theta - s_\alpha^2 \left(c_\theta + s_\theta \frac{v}{w}\right) f(R,\epsilon)\right] \,,\nonumber \\
h_{125}\,Z^\mu_D \, Z^D_{\mu} &\longrightarrow \frac{2 m_Z^2}{v} \left[c_\theta R - c_\alpha^2 \left(c_\theta + s_\theta \frac{v}{w}\right) f(R,\epsilon)\right] \,,\nonumber \\
h_{125}\,Z^\mu_D \, Z_{\mu}  &\longrightarrow \frac{4\,m_Z^2}{v} \left[ c_\alpha s_\alpha \left(c_\theta + s_\theta \frac{v}{w}\right) f(R,\epsilon) \right]  \, , \\
h_{D}\,Z^\mu Z_{\mu} & \longrightarrow  \frac{2 m_Z^2}{v} \left[s_\theta - s_\alpha^2 \left(s_\theta - c_\theta \frac{v}{w}\right) f(R,\epsilon)\right] \,,\nonumber\\
h_{D}\,Z^\mu_D \, Z^D_{\mu}  & \longrightarrow \frac{2 m_Z^2}{v} \left[s_\theta R - c_\alpha^2 \left(s _\theta - c_\theta \frac{v}{w}\right) f(R,\epsilon)\right]  \,,\nonumber \\
h_{D}\,Z^\mu_D \, Z_{\mu} & \longrightarrow \frac{4\,m_Z^2}{v} \left[ c_\alpha s_\alpha \left(s_\theta - c_\theta \frac{v}{w}\right) f(R,\epsilon) \right] \,,\nonumber
\end{align}
with $f(R,\epsilon) = c_\alpha^2 \left( R -\delta^2 \right) + s_\alpha^2 \left(1 -\delta^2 R \right)$ and $f(R,\epsilon) \times m_Z^2 \equiv M_{X_0}^2$.

The mixing in the scalar sector modifies the couplings of
$h_{125}$ with respect to their SM values, and is thus constrained by Higgs signal strength measurements performed by the ATLAS~\cite{ATLAS:2019slw} and CMS~\cite{CMS:2018uag} collaborations, yielding the 95\% C.L.~limit~\cite{Biekotter:2022ckj}
\begin{equation}
    (1 - {\rm BR}_{h_{125} \to {\rm BSM}}) \, c_\theta^2 \geq 0.923\, ,
    \label{eq:Higgs_signal_strengths}
\end{equation}
with ${\rm BR}_{h_{125} \to {\rm BSM}}$ the 125 GeV Higgs boson branching fraction into non-SM final states (e.g. $h_{125} \to Z_D Z_D$, $h_{125} \to Z_D Z$ or $h_{125} \to h_D h_D$). For ${\rm BR}_{h_{125} \to {\rm BSM}} = 0$ the above limit yields $s_{\theta} < 0.277$ (a slightly stronger bound is obtained in Ref.~\cite{Fernandez-Martinez:2022stj}).

The singlet-like dark Higgs boson $h_D$ can be either heavier or lighter than the 125 GeV Higgs boson $h_{125}$, leading in both cases to a rich collider phenomenology.  A heavy state $h_D$ can, e.g., be searched for in $W W$, $Z Z$ or di-Higgs final states (see, e.g., Refs.~\cite{ATLAS:2018rnh,ATLAS:2018sbw,CMS:2018kaz,No:2013wsa,Buttazzo:2015bka,Huang:2017jws}). At the same time, a light Higgs boson $h_D$ which decays into SM states is subject to strong constraints from direct searches at the LHC and LEP~\cite{LEP:2003ing}. The possibility of an exotic decay of the 125 GeV Higgs boson into singlet-like Higgs bosons $h_{125} \to h_D h_D$, occurring for $m_{h_D} < m_{h_{125}}/2 \simeq 62.5$~GeV, is also constrained both indirectly via \cref{eq:Higgs_signal_strengths} and through direct searches for exotic Higgs decays (see Ref.~\cite{Cepeda:2021rql} for a recent review). Finally, there exist searches at LEP for $e^+ e^- \to Z\, h_D$ which target an invisibly decaying $h_D$~\cite{ALEPH:2001roc,DELPHI:2003azm,L3:2004svb,OPAL:2007qwz} (which could occur via $h_D \to Z_D Z_D$) and constrain the mixing $s_\theta$, placing a stringent limit $s_\theta < 0.1$ for $m_{h_D} < 40$~GeV (see e.g. Ref.~\cite{Fernandez-Martinez:2022stj}). In any case, we refrain from a more detailed discussion here and refer the reader to Refs.~\cite{Falkowski:2015iwa,Robens:2015gla,Assamagan:2016azc,Arcadi:2019lka,Bauer:2020nld,Bhattiprolu:2022ycf} for recent phenomenological reviews on the singlet scalar extension of the SM. Recent phenomenological analyses of the DAHM or closely related models (with some SM fermions charged under the $U(1)_X$ gauge symmetry) can also be found in Refs.~\cite{Foguel:2022unm,Duerr:2017uap,Ferber:2023iso}.

Before continuing to the next section, we stress an alternative possibility to obtain a massive dark photon $Z_D$
via a  St\"uckelberg mechanism, where the longitudinal degree of freedom (d.o.f.) of the dark photon is provided by a new pseudo-scalar $\sigma$, with the relevant part of the Lagrangian given by 
\begin{multline}
    \mathcal{L} = -\frac{\epsilon}{2} \hat{X}_{\mu\nu}\hat{B}^{\mu\nu} -\frac{1}{4} \hat{X}_{\mu\nu} \hat{X}^{\mu\nu} + \frac{1}{2} M_{X_0}^2\left(\hat{X}_\mu - \frac{1}{M_{X_0}} \partial_
    \mu \sigma\right)^2 
    -\frac{1}{2\xi} \left(\partial_\mu\hat X^\mu +\xi\, M_{X_0}\,\sigma\right)^2 \,,
\end{multline}
with $\xi$ being the gauge fixing parameter.
This St\"uckelberg mechanism yields a massive dark photon, but without an extra scalar d.o.f.~$h_D$ in the physical spectrum. This effectively amounts to setting $s_\theta = 0$ in the interactions~\eqref{eq:DAHM_interactions} for the DAHM and considering only the SM Higgs $h_{125} = h$ as a physical scalar particle in the spectrum.

\subsection{Adding a dark fermion as Dark Matter}
\label{sec:darkfermion}

Finally, let us specify the matter content of the dark gauge current, $j_\mu^X$, introduced in \cref{eq:loop_lag}. In this work, we consider a Dirac fermion, $\chi$, that transforms as a singlet under the SM gauge group and carries charge $q_\chi$ under the $U(1)_X$ symmetry, such that
\begin{equation}
\label{dark_current}
    j_\mu^X = q_\chi\, \bar \chi \, \gamma_\mu\, \chi \,.
\end{equation}
We will set $q_\chi=1$ for simplicity throughout the remainder of this article. The interactions between the gauge fields and the fermionic currents from \cref{eq:loop_lag_diag2} can be written in terms of the physical fields $A^\mu$, $Z^\mu$, $Z_D^\mu$ after rotating to the mass eigenstates via~\cref{eq:phys_diag},
\begin{align}
\label{eq:fermion_currents}
    \mathcal{L} \supset &- A^{\mu} (e\, j^\mathrm{EM}_\mu) - Z^\mu \left[ g_Z\, j^Z_\mu \left(c_\alpha - s_\alpha \delta \right) + e\, j^\mathrm{EM}_\mu (s_\alpha \delta/t_W) - g_D \,j_\mu^X \, s_\alpha \right] \nonumber \\
    &- Z_D^\mu \left[ g_D \,j_\mu^X \,  c_\alpha + g_Z\, j^Z_\mu (s_\alpha + c_\alpha \delta ) - e\, j^\mathrm{EM}_\mu (c_\alpha \delta/t_W) \right] \, .
\end{align}

Given that $\chi$ is a Dirac fermion, a Yukawa term $S \,  \bar \chi \chi$ is forbidden in $\mathcal{L}$ as it is not $U(1)_X$ gauge invariant.\footnote{We note that if $\chi$ were instead a Majorana fermion, the Yukawa term $S \,  \bar \chi^c \chi$ would be gauge-invariant (for appropriate choices of the $U(1)_X$ charges for $S$ and $\chi$), and yield a scalar portal to DM within the DAHM scenario.
In this case, both vector and scalar portal to DM could play a relevant role in the DM phenomenology. We will discuss this possibility elsewhere.} Conversely, the mass term $m_\chi \, \bar \chi \chi$ is gauge invariant, and we add it to the Lagrangian of our model (with $m_{\chi}$ a free parameter).

The singlet fermion $\chi$ can be a viable DM candidate. In this minimal setup, the only interactions between DM and the SM proceed via the vector portal (recall that the scalar portal is forbidden by $U(1)_X$ gauge invariance). This DM scenario has been well-studied in the light mediator regime with $m_{Z_D} \lesssim 10$~GeV, where the invisible dark photon can be directly searched for at $B$-factories, beam dump and fixed target experiments (see, e.g., \cite{Gori:2022vri,Antel:2023hkf} for recent reviews). However, the GeV region with 10 GeV $ \lesssim m_{Z_D} < m_Z$ is more challenging to probe via collider experiments, being above the reach of $B$-meson factories. In this section, we discuss the production of the thermal relic abundance of $\chi$ coupled to such a GeV-scale dark photon mediator.

We consider a freeze-out scenario, where the DM was in thermal equilibrium with the SM bath in the early universe.
To compute the DM relic abundance produced during thermal freeze-out, we have to solve the Boltzmann equation
\begin{equation}\label{eq:BE}
    \frac{d Y}{d x}=\frac{s\langle\sigma v\rangle}{H x}\left[1+\frac{1}{3} \frac{d\left(\ln g_s\right)}{d(\ln T)}\right]\left(Y_{e q}^2-Y^2\right),
\end{equation}
tracking the time evolution of the DM number density per comoving volume, $Y = n_\chi/s$. Here, $s$ denotes the entropy density, $H$ is the Hubble parameter, $x=m_\chi/T$ and $g_s$ are the number of relativistic degrees of freedom contributing to the entropy density.
The DM abundance freezes out once $\chi$ goes out of chemical equilibrium with the SM particle bath. This is typically the case once the Hubble expansion rate $H$ exceeds the DM annihilation rate, $n_\chi \langle\sigma v\rangle$.
The annihilation rate is driven by the thermally averaged annihilation cross-section in the early universe, which we can determine numerically from~\cite{Gondolo:1990dk,Griest:1990kh},
\begin{equation}\label{eq:therm_av}
    \langle \sigma v \rangle_\text{CM} = \frac{x}{2\ [K_1^2(x)+K_2^2(x)]} \,\int_{2}^\infty\dif z \ \sigma\left(z^2m_\chi^2\right)\, (z^2-4)\, z^2\,K_1(zx) \,,
\end{equation}
where $\sigma(s)$ is the annihilation cross-section,  $z=\sqrt{s}/m_\chi$  and
$K_n(x)$ denote the modified Bessel functions of the second kind. Following Ref.~\cite{Steigman:2012nb}, one can solve~\cref{eq:BE} approximately analytically and from the resulting frozen-out yield  $Y_f$ one can compute the DM relic abundance as $\Omega h^2_\chi={m_\chi Y_f s_0 h^2}/{\rho_{c}}$, with $s_0$ denoting the entropy density today and the critical density $\rho_{c}=3 H_0^2/8 \pi G$ in terms of the Hubble rate $H_0$ and gravitational constant $G$. A more complete computation that does not rely on the assumption that DM is kept in kinetic equilibrium with the SM heat bath during the entire chemical decoupling process can be found in Refs.~\cite{Binder:2017rgn,Binder:2021bmg}. Note also that while the formalism of Ref.~\cite{Steigman:2012nb} can be straightforwardly extended to higher partial wave ($p$, $d$, etc.), it was developed primarily for $s$-wave dominated annihilation processes. While in our numerical computations we have always retained the full thermal dependence of the cross section, we have explicitly checked that the DM annihilation processes in our model are $s$-wave dominated everywhere.

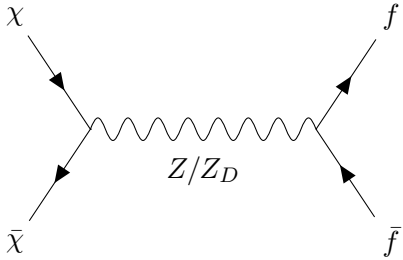
\begin{figure} [!t]
    \centering
    \begin{tikzpicture}
        \begin{feynman}
            \vertex (p1) at (-2.5, 1.5) {$\chi$};
            \vertex (p2) at (-2.5, -1.5) {$\bar\chi$};
            \vertex (p3) at (2.5, 1.5) {$f$};
            \vertex (p4) at (2.5, -1.5) {$\bar f$};
            \vertex [crossed dot, minimum size=0pt](v1) at (-1.5, 0) {};
            \vertex [crossed dot, minimum size=0pt](v2) at (1.5, 0) {};

            \diagram* {
                (p1) -- [fermion] (v1), 
                (v1) -- [yourboson, edge label'=$Z/Z_D$, inner sep=10pt] (v2),
                (v2) -- [fermion] (p3),
                (v1) -- [fermion] (p2),
                (p4) -- [fermion] (v2)
            };
        \end{feynman}
    \end{tikzpicture}
    \caption{\label{fig:diagrams_ann} Dominant annihilation channel of the DM candidate $\chi$ in the regime $m_{\chi}\lesssim m_{Z_D}/2$.}
\end{figure}

In this work, we are interested in the regime in which the decay of the dark photon can proceed invisibly into pairs of DM particles ($ m_{\chi} \lesssim m_{Z_D}/2$). In the resonance region, strong direct detection constraints can be evaded (c.f.~\cref{sec:dd}). In this regime, the DM annihilations that set the relic density in the early universe proceed via the $2\to 2$ processes $\chi\,\bar\chi \to f \bar f$ illustrated in~\cref{fig:diagrams_ann}, with $f$ any electrically charged SM fermion. We stress that above the resonance, $m_{\chi} > m_{Z_D}/2$, additional processes may contribute to the freeze-out relic density. In particular, it has been shown that above the resonance, in the regime $2\,m_\chi\gtrsim m_{Z_D}\gtrsim  m_\chi$, the $3\to2$ process $\chi\bar \chi \chi \to \chi\, Z_{D}$ can be very relevant~\cite{Cline:2017tka,Fitzpatrick:2020vba}. Moreover, there is the forbidden $2\to 2$ process $\chi\bar\chi\to Z_{D} Z_{D}$, which can contribute significantly to the relic density if there is an exponential hierarchy between the weak scale and the DM mass~\cite{DAgnolo:2015ujb}, which is not satisfied in this analysis. In addition, far above the resonance, the annihilation channels $\chi\,\chi \to Z_D\, Z_D$ and $\chi\,\chi \to Z_D\, h_D$ can become kinematically allowed. As we focus on the $ m_{\chi} \lesssim m_{Z_D}/2$ mass range, all these processes do not play a role in our analysis.

While in our numerical calculation we rely on the full annihilation cross-section, including $Z-Z_D$ interference, we want to comment on the parametric scaling of the dominant $s$-channel cross-section of the $\bar\chi \chi \to \bar f f$ process. Away from the $Z$ pole, $m_{Z_D}<m_Z$, the DM annihilation process is dominated by exchange of a $Z_D$ both due to its lighter mass and parametrically larger couplings. There, the annihilation cross-section scales as,
\begin{equation}
    \sigma_\mathrm{ann} \propto \frac{\epsilon^2\,\alpha\,\alpha_D}{(s-m_{Z_D}^2)^2 + m_{Z_D}^2 \Gamma_{Z_D}^2 }\,,
\end{equation}
with $\alpha=e^2/4\pi$, $\alpha_D=g_D^2/4\pi$, $\sqrt{s}$ the centre of mass energy of the annihilation and $\Gamma_{Z_D}$ the dark photon decay width (see section \ref{section:dilepton}). From this scaling, we can see that the produced relic abundance is controlled by the product $\epsilon^2 \, \alpha_D$ of the kinetic mixing parameter and dark sector coupling. Hence, bounds on the visible sector couplings $\epsilon$ can generally be avoided by increasing the dark sector coupling, $\alpha_D$. Furthermore, due to the $s$-channel resonance in the $Z_D$ propagator, the annihilation cross-section can be significantly enhanced close to the resonance. Since during freeze-out in the early universe the DM $\chi$ can still have considerable kinetic energy, one can observe a thermal broadening of this resonance enhancement leading to a broad resonance peak in the relic density~\cite{Guo:2009aj} (c.f.~\cref{fig:DDandRD}).

\section{Constraints}
\label{sec:constraints}

In this section, we review the most stringent constraints on the \dmmodel, coming from direct and indirect DM searches, as well as collider experiments.

\subsection{Direct Detection}
\label{sec:dd}

\begin{figure} [!t]
    \centering
    \begin{tikzpicture}
        \begin{feynman}
            \vertex (p1) at (-1.8, 1.8) {$\chi$};
            \vertex (p2) at (-1.8, -1.8) {$N$};
            \vertex (p3) at (1.8, 1.8) {$\chi$};
            \vertex (p4) at (1.8, -1.8) {$N$};
            \vertex [crossed dot, minimum size=0pt](v1) at (0, 1) {};
            \vertex [dot, minimum size=7pt](v2) at (0, -1) {};

            \diagram* {
                (p1) -- [fermion] (v1), 
                (v1) -- [yourboson2, edge label'=$Z$, inner sep = 10] (v2),
                (p2) -- [fermion] (v2),
                (v1) -- [fermion] (p3),
                (v2) -- [fermion] (p4)
            };
        \end{feynman}
    \end{tikzpicture}
    \quad
    \begin{tikzpicture}
        \begin{feynman}
            \vertex (p1) at (-1.8, 1.8) {$\chi$};
            \vertex (p2) at (-1.8, -1.8) {$N$};
            \vertex (p3) at (1.8, 1.8) {$\chi$};
            \vertex (p4) at (1.8, -1.8) {$N$};
            \vertex [crossed dot, minimum size=0pt](v1) at (0, 1) {};
            \vertex [dot, minimum size=7pt](v2) at (0, -1) {};

            \diagram* {
                (p1) -- [fermion] (v1), 
                (v1) -- [yourboson2, edge label'=$Z_D$, inner sep = 10] (v2),
                (p2) -- [fermion] (v2),
                (v1) -- [fermion] (p3),
                (v2) -- [fermion] (p4)
            };
        \end{feynman}
    \end{tikzpicture}
    \caption{\label{fig:diagrams1} Relevant diagrams for the elastic scattering of DM off nuclei, $N$.}
\end{figure}
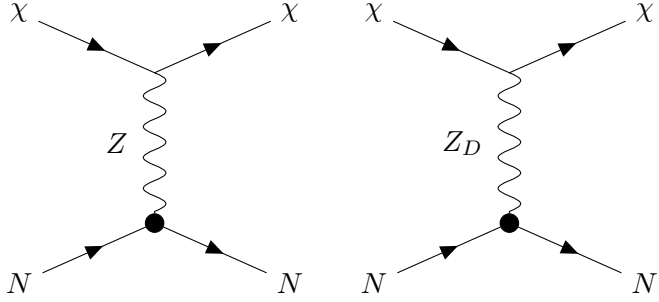

In this model, the fermionic DM candidate $\chi$ interacts with SM particles primarily through the exchange of the dark photon and the SM $Z$ boson, both of which mediate elastic scattering processes through the diagrams of \cref{fig:diagrams1}. The resulting scattering cross-section therefore depends on the kinetic mixing and the dark gauge coupling as $\epsilon^2 \, \alpha_D$. Since the DM is a Dirac particle, the absence of scalar-mediated interactions further simplifies the phenomenology. 

The expected rate of DM-induced nuclear recoils in a direct detection experiment is obtained by integrating the DM-nucleus differential scattering cross-section, $d\sigma_{\chi T}/{dE_R}$, multiplied by the DM velocity distribution function in the halo, $f(\vec{v})$ (see e.g., Ref.~\cite{Cerdeno:2010jj}),
\begin{equation} \label{eq:DDrate}
    \frac{dR}{dE_R}=\frac{\xi\rho_0}{m_Tm_\chi}\ \epsilon(E_R)\ \int dE'_R\ Gauss(E'_R,E_R)\int_{v_\text{min}}d^3v\,v\,f(\vec{v})\,\frac{d\sigma_{\chi T}}{dE'_R}\, ,
\end{equation}
where $m_T$ is the target nuclei mass, $v_\text{min} = \sqrt{{m_T E_R}/{2 \mu_T^2}}$ is the minimum dark matter velocity required to produce a nuclear recoil with energy $E_R$, and $\mu_T$ is the dark matter–nucleus reduced mass. We take into account the specific features of each experiment through the energy-dependent functions of efficiency $\epsilon(E_R)$ and resolution $Gauss(E'_R,E_R)$ (incorporated as a Gaussian smearing). We follow the generic prescription of Ref.~\cite{Baxter:2021pqo} to fix the parameters of the DM halo to standard values, and we take $\rho_0=0.3$~GeV/cm$^2$ for the total DM local density. The local density of $\chi$, taking into account the dilution factor, is therefore $\xi\rho_0$. Notice from this expression that DD experiments can only set constraints on the product $\xi d\sigma_{\chi T}/{dE'_R}$ (or $\xi \sigma_{SI}$ as we will see below when we introduce the spin-independent contribution). Our numerical calculation is performed with the RAPIDD package \cite{Cerdeno:2018bty}.

The main contribution to the DM-nucleus scattering in this model is the usual spin-independent interaction. Although one can use the language of non-relativistic effective field theory to show that there are also contributions to velocity and momentum dependent operators ${\cal O}_7$ and ${\cal O}_9$ in the prescription of Ref.~\cite{Fitzpatrick:2012ix}, these are extremely small and can be safely ignored. Thus, the differential cross-section can be written as \cite{Evans:2017kti}:
\begin{equation}
        \frac{d\sigma_{\chi T}^{SI}}{dE_R}=\frac{m_T}{2\pi v^2}\ g_D^2\,\epsilon^2\,A^2\,F^2(E_R)\left|\frac{f_n^{(Z_D)}}{m_{Z_D}^2}+\frac{f_n^{(Z)}s_W}{m_Z^2-m_{Z_D}^2}\right|^2\, ,
\end{equation}
where $A$ is the mass number of the target nucleus, $F^2(E_R)$ is the Helm nuclear form factor \cite{PhysRev.104.1466,Lewin:1995rx} as a function of the recoil energy $E_R$. The effective DM couplings to protons and neutrons, $f_n^{(Z_D)}$ and $f_n^{(Z)}$ are given by:
\begin{equation}
    f_n^{(X)}=\frac{1}{A}\left(Zf_{p,X}+(A-Z)f_{n,X}\right)\, ,
\end{equation}
where $Z$ is the atomic number of the target nucleus, $f_{p,X}=2g_{u,X}+g_{d,X}$, $f_{n,X}=g_{u,X}+2g_{d,X}$ 
and $g_{u,X}$  and $g_{d,X}$ are the couplings of the boson $X=Z,Z_D$ with up and down quarks that can be obtained from \cref{eq:fermion_currents}. 
Note that for $Z_D$ masses of $m_{Z_D}<m_{Z}$ the interactions of $Z_D$ in~\cref{eq:fermion_currents} are to very good approximation only with the electromagnetic current. This effectively suppresses any interactions with neutrons.

Typically, the exclusion lines derived from direct detection data assume that the DM couplings to protons and neutrons are equal, $\sigma_{SI}^n = \sigma_{SI}^p \equiv \sigma_{SI}^\text{iso}$. Since this is not the case in this model, we need to re-cast these results accordingly:

\begin{equation}
    \sigma_{SI}^p =\frac{A^2}{Z^2}\sigma_{SI}^\text{iso}\, .
    \label{eq:sigmap}
\end{equation}

\begin{figure}[!t]
\begin{centering}
    \includegraphics[width=0.9\textwidth]{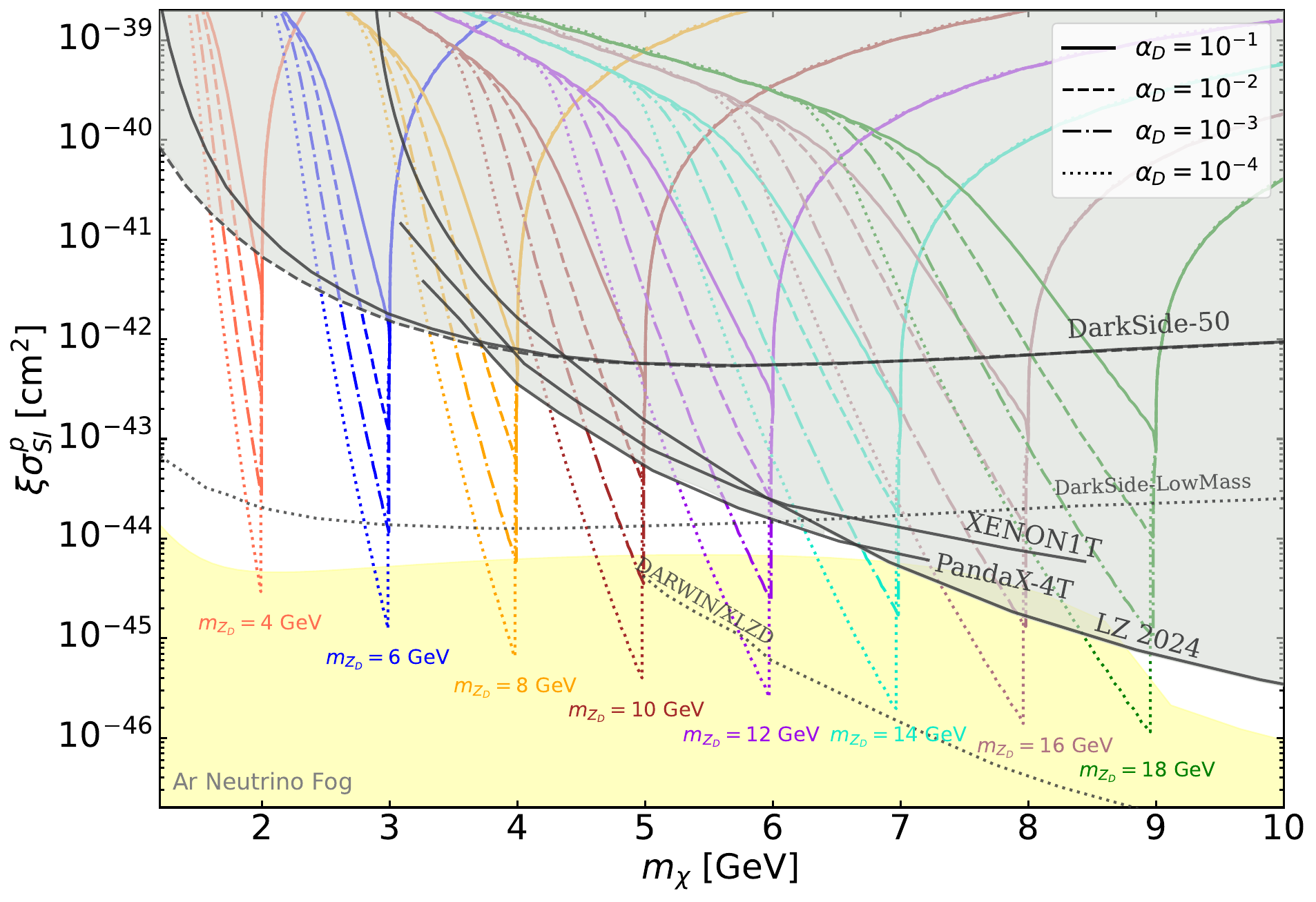}
    \caption{
    Coloured lines represent the theoretical predictions for $\xi\sigma_{SI}^p$ s a function of the DM mass for several values of the dark gauge coupling (represented with different line styles) and different dark photon masses (in different colours). As explained in the main text, $\xi\sigma_{SI}^p$ is independent of $\epsilon$, and each of these lines represent all values of the relic abundance (or $\epsilon$).  Current experimental constraints from DarkSide-50 \cite{DarkSide-50:2022qzh}, XENON1T \cite{XENON:2020gfr}, PandaX-4T \cite{PandaX-4T:2021bab} and LUX-ZEPLIN (LZ) \cite{LZ:2024zvo} are shown as a gray area. We also show as dotted gray lines projections from future DD experiments DarkSide-LowMass \cite{GlobalArgonDarkMatter:2022ppc} and DARWIN/XLZD \cite{DARWIN:2016hyl,Aalbers:2022dzr,XLZD:2024nsu}. The pale yellow region corresponds to the neutrino floor for an argon target \cite{PhysRevLett.127.251802}.}
    \label{fig:DDandRD}
\end{centering}
\end{figure}

For WIMP masses below roughly $10$~GeV, the sensitivity of xenon-based detectors, such as XENON1T, LUX-ZEPLIN, and PandaX, begins to deteriorate, owing to the heavy xenon nucleus’s reduced recoil energy. In this low-mass regime, experiments employing lighter target nuclei, such as the argon-based DarkSide program, silicon-and-germanium detectors in SuperCDMS, or fluorine-rich bubble chambers like PICO, can achieve lower energy thresholds and thus set stronger exclusion limits.

In \cref{fig:DDandRD}, we show the theoretical predictions for $\xi\sigma_{SI}^p$ as a function of the DM mass, for different dark photon masses $m_{Z_D}$ (in different colours) and dark gauge couplings $\alpha_D$ (in different line styles). Notice that since the scattering cross-section scales as $\sigma_{SI}^p\propto\epsilon^2$, while the dilution factor scales as $\xi\sim 1/\langle\sigma v\rangle\propto \epsilon^{-2}$, the theoretical predictions for $\xi\sigma_{SI}^p$ do not change when we consider cases with subdominant $\chi$. In other words, DD bounds are {\em independent of the value of $\xi$ (or $\epsilon$)}: if a given range of $m_\chi$ is allowed (excluded) for $\xi=1$, it remains allowed (excluded) for any value $\epsilon$. This observation is crucial when we combine DD constraints with ID and collider limits in \cref{sec:res}.

\begin{figure}[!t]
\begin{centering}
    \includegraphics[width=0.9\textwidth]{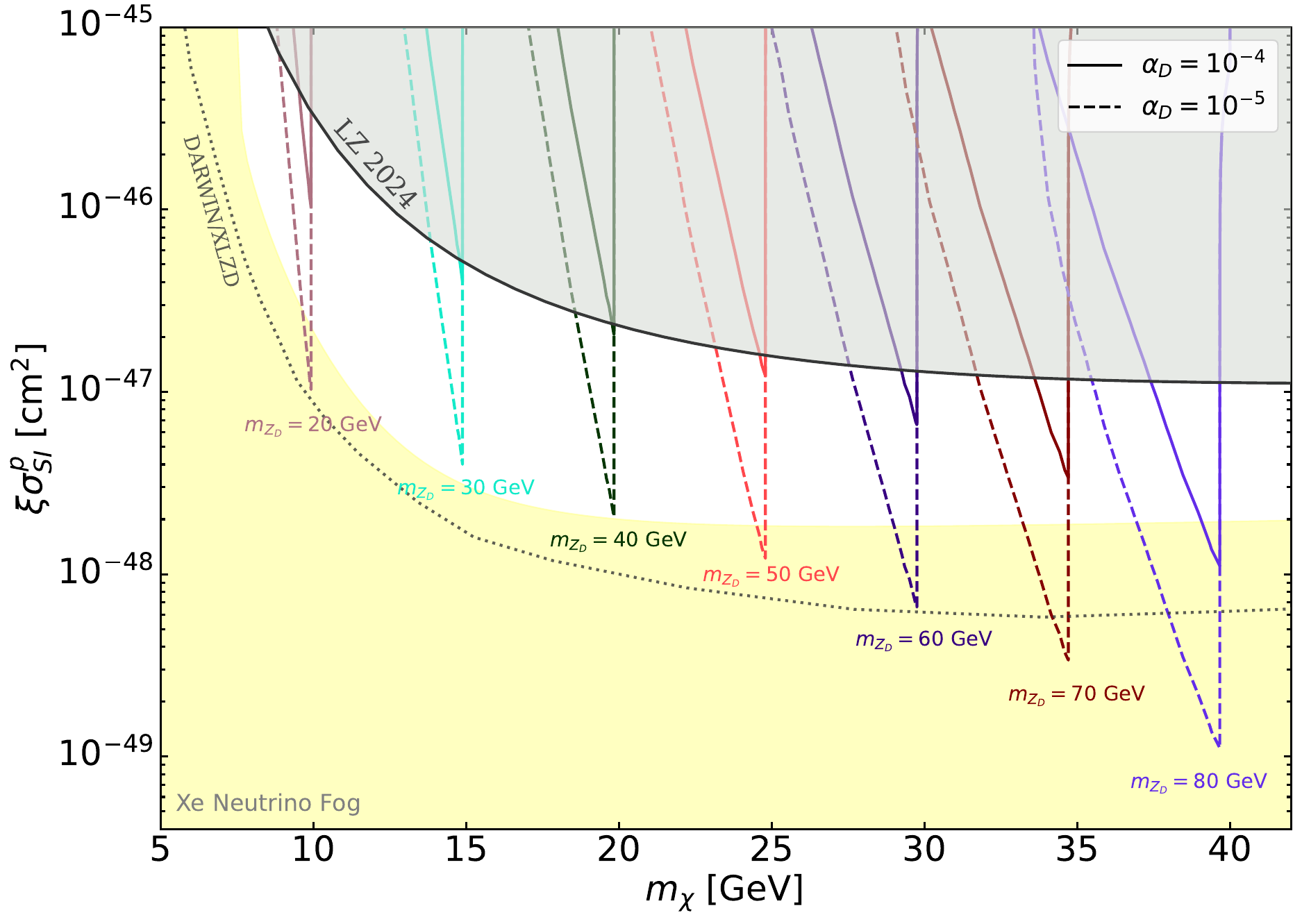} 
    \caption{
    The same as \cref{fig:DDandRD}, but for dark photon masses above $20$~GeV. The pale yellow region corresponds here to the neutrino floor for a xenon target \cite{PhysRevLett.127.251802}.}
    \label{fig:DDandRD_HighMass}
\end{centering}
\end{figure}

We also display the leading DD constraints on DM-proton scattering cross-section. For DarkSide-50 \cite{DarkSide-50:2022qzh}, XENON1T \cite{XENON:2020gfr} and PandaX-4T \cite{PandaX-4T:2021bab} we have re-scaled the published bounds according to~\cref{eq:sigmap}. For LUX-ZEPLIN (LZ) \cite{LZ:2024zvo} we have re-computed the expected rates using RAPIDD and we have done our own likelihood analysis in order to extend the published bound to masses below $9$~GeV. 
It can be seen that, away from the resonances, the theoretical predictions for $\xi\sigma_{SI}^p$ are ruled out by direct detection constraints for all values of the dark gauge coupling shown here. However, in the vicinity of the resonances, there is a window in which this model is still not excluded, specially for small values of the dark photon mass. As the dark photon mass increases, direct detection bounds become more stringent due to the enhanced sensitivity of xenon-based experiments. Even in those cases the limits can be avoided and the model can still be viable if the dark gauge coupling is low enough. In the plot, we also show projections for the future DD argon and xenon experiments DarkSide-LowMass \cite{GlobalArgonDarkMatter:2022ppc} and DARWIN/XLZD \cite{DARWIN:2016hyl,Aalbers:2022dzr,XLZD:2024nsu}, which will push the available windows for this model to smaller values of the gauge coupling. To explore that region, DD experiments will enter the so-called neutrino floor \cite{PhysRevLett.127.251802}, where DM signals become increasingly difficult to distinguish from the irreducible background produced by neutrino interactions, due to their similar spectral characteristics. We show this region for argon-target experiments (although it is similar for xenon ones) in pale yellow.

In \cref{fig:DDandRD_HighMass}, we extend our analysis for dark photon masses above $20$~GeV and DM masses above $10$~GeV. In this range, to avoid the DD constraints (dominated by LZ results), the dark gauge coupling has to be further reduced to $\alpha_D\sim10^{-5}$. The theoretical predictions populate the available area before entering the neutrino floor. The future DARWIN/XLZD experiment, which will clearly reach this region with neutrinos dominating the background, will be able to probe the allowed windows for the model in this mass regime.

\subsection{Indirect Detection}
\label{sec:id}

In this \dmmodel scenario, DM particles can annihilate into SM particles via an $s$-channel $Z_D$ and $Z$ exchange. This can lead to various interesting signals that can be searched for in indirect detection experiments, like the observation of the CMB or the search for gamma ray emission from dwarf galaxies.

\subsubsection{Gamma rays from dwarf spheroidal galaxies}

In recent years, there has been an ever growing catalogue of identified  dwarf spheroidal galaxies (dSphs) discovered e.g.~in the Dark Energy Survey (DES)~\cite{DES:2019vzn}. The Fermi-Large Area Telescope (Fermi-LAT) has now observed Milky Way satellite dSphs  for over 15 years and released updated data in the 4FGL-DR3 source catalogue~\cite{Fermi-LAT:2022byn}. 
These dSphs are prime targets for studying the micro-physical properties of DM since they are some of the most DM-rich environments in our universe~\cite{Battaglia:2013wqa,Strigari:2018utn}. Therefore, some of the leading constraints on DM annihilations in the universe stem from gamma ray observations of dSphs. The current leading bound is provided for by the Fermi-LAT legacy analysis presented in Ref.~\cite{McDaniel:2023bju}. In this paper, we will reproduce their analysis closely and follow their statistical analysis to derive constraints on the annihilation into gamma rays on our DM model.

The expected differential flux of gamma rays produced in DM annihilations in dSphs in a solid angle $\Delta\Omega$ is given by 
\begin{equation}\label{eq:dflux}
    \frac{d \phi}{dE} (\Delta \Omega) = \frac{J}{4\pi} \sum_f\frac{\langle \sigma v \rangle_f}{2 \, m_\chi^2} \, \frac{d N^f_\gamma}{dE} \,,
\end{equation}
with the $J$-factor being the line-of-sight integral across the DM distribution within a solid angle $\Delta \Omega$,
\begin{equation}
    J= \int_{\Delta \Omega}d\Omega\int d\ell \, \rho_\chi(r)^2\,.
\end{equation}
In fact, the $J$-factor depends sensitively on the galactic density profile of $\chi$. For consistency, we consider that the galactic abundance of $\chi$ is the same fraction of the total DM density as dictated by the cosmological production, i.e., $\rho_\mathrm{\chi}=\xi\, \rho_\mathrm{CDM}$. Notice that, in practice, this means rescaling the $J$-factor or equivalently the flux in~\cref{eq:dflux} by the square of the dilution parameter, $\xi^2$.

\begin{figure}[!t]
\begin{centering}
    \includegraphics[width=0.49\textwidth]{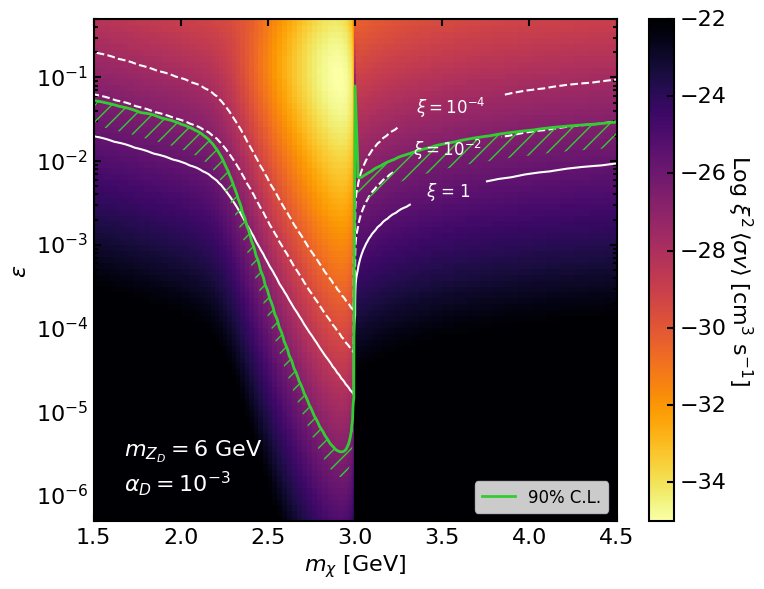} %
    \includegraphics[width=0.49\textwidth]{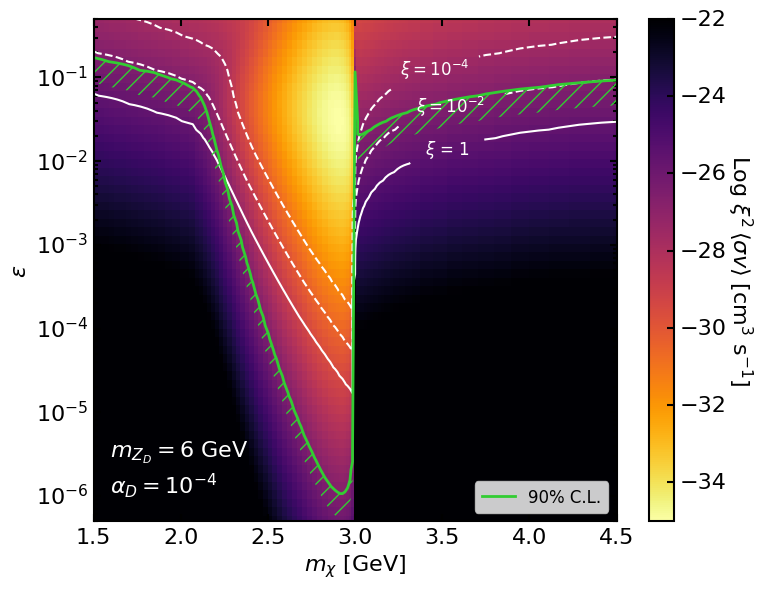} \\
    \includegraphics[width=0.49\textwidth]{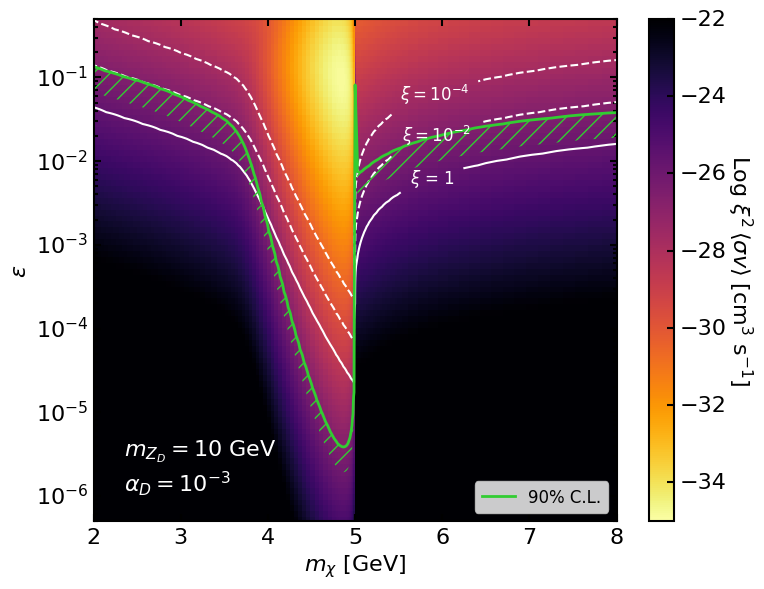} %
    \includegraphics[width=0.49\textwidth]{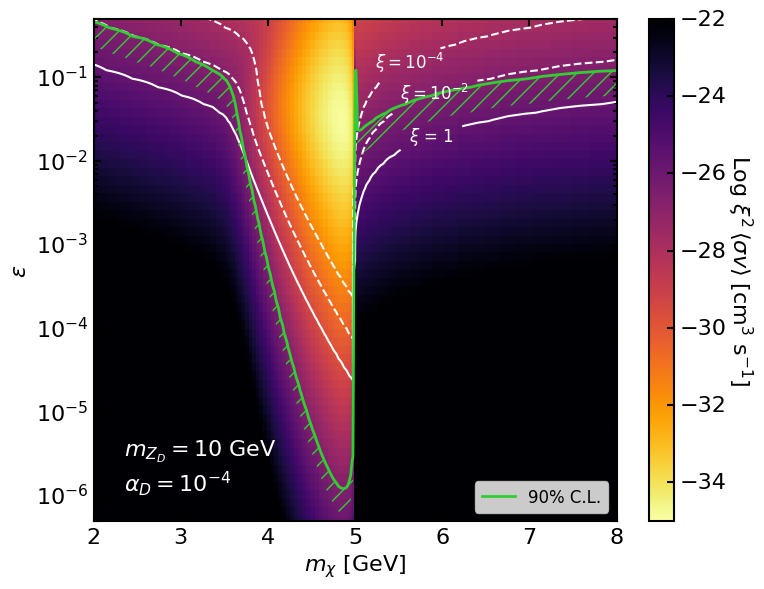}
    \caption{Limits from gamma ray flux measurements of dSphs with the Fermi-LAT legacy data set~\cite{McDaniel:2023bju} for masses of $m_{Z_D} = 6$~GeV (top) and $m_{Z_D} = 10$~GeV (bottom)  and couplings of $\alpha_D=10^{-3}$ (left) and $\alpha_D=10^{-4}$ (right). The green line denotes the 90\% C.L.~exclusion limit derived from the Fermi-LAT likelihood. Note that all parameter space \textit{below} the green line is excluded.  The white solid and dashed lines represent the relative abundance of $\chi$ for $\xi=1, 10^{2}$ and $10^{-4}$, respectively. We can see that the correct relic abundance ($\xi=1$) is excluded everywhere except for the resonance region.  }
\label{fig:gamma_ray}
\end{centering}
\end{figure}

The Fermi-LAT legacy analysis of Ref.~\cite{McDaniel:2023bju} provides the single-source, binned likelihood profiles, $\mathcal{L}_i(d\phi/dE, E)$, as a function of the differential gamma ray flux $d\phi/dE$ and energy $E$ for different sets of dSPhs source catalogues. We can obtain the full model likelihood by computing the differential gamma ray fluxes $d\phi/dE$ according to~\cref{eq:dflux} and multiplying all the single-source likelihoods for the observed dSphs. For each source, the uncertainty on the observed $J$-factor is taken into account as a nuisance parameter,
\begin{equation}
    \mathcal{L}_i(d\phi/dE, E, J_{obs}) =  
    \frac{\mathcal{L}_i(d\phi/dE, E)}{\ln (10) \sqrt{2 \pi \sigma_J} J_{o b s}} \ \exp \left[-\left(\frac{\log _{10}(J)-\log _{10}\left(J_{o b s}\right)}{\sqrt{2} \sigma_J}\right)^2\right]\,,
\end{equation}
 and to obtain the final likelihood we profile over the $J$-factors. In our analysis we have considered the catalogue of 30 dSphs labelled `\textit{measured}' in Ref.~\cite{McDaniel:2023bju}, for which the $J$-factors have been measured.

From the model likelihoods, we build a test statistic according to the profile log-likelihood ratio~\cite{Feldman:1997qc},
 \begin{equation}
     \log \lambda(\theta) = \log \frac{\mathcal{L}(\theta)}{\mathcal{L}(\theta_0)} \,,
 \end{equation}
where $\theta$ denotes our model parameters and $\theta_0$ are the model parameters that maximise the likelihood.  Under random fluctuations in the data, the test statistic $-2 \log \lambda(\theta)$ is asymptotically distributed as a $\chi^2$ function with the number of degrees of$d$ equal to the number of tested parameters $\theta$. Hence, for a fixed DM mass we leave the kinetic mixing $\epsilon$ freely floating. We can thus reject the model at the $1-p$ level if
\begin{equation}
    1-F_{\chi^2}(d=1, -2 \log\lambda(\theta)) <p \,,
\end{equation}
with $F_{\chi^2}$ being the cumulative ${\chi^2}$ distribution with $d=1$ d.o.f.  We compute the one-sided 90\% confidence-level (\cl) exclusion limit from $-2 \log\lambda(\theta) > 2.71$.

In~\cref{fig:gamma_ray}, we show the resulting limits on the $(\epsilon,\, m_{\chi})$ parameter space for four different benchmark scenarios, with $m_{Z_D}=6\,(10)$~GeV in the top (bottom) row and $\alpha_D=10^{-3}\,(10^{-4})$ in the left (right) column, respectively. The green line represents the $90\%$ \cl exclusion limit (the area underneath is ruled out). The colour map represents the corresponding value of $\xi^2\langle\sigma v\rangle$, which increases towards smaller values of $\epsilon$ due to the rapid increase of the relic abundance and the $\xi^2$ factor. For reference, we show the DM abundance curves for three different values of $\xi\in\{1, 10^{-2}, 10^{-4}\}$. Therefore, as the kinetic mixing increases and the relic abundance of $\chi$ hence becomes smaller, the gamma ray flux from DM pair annihilation decreases as $\epsilon^{-2}$. As a consequence, and contrary to how the Fermi-LAT bound is often displayed when the dilution factor is not considered, we obtain a {\em lower} bound on the kinetic mixing $\epsilon$ (and on $\xi^2 \langle\sigma v\rangle$). Notice from \cref{fig:gamma_ray} that this lower bound excludes the correct relic density ($\xi=1$) line for DM masses outside the resonance.

\subsubsection{CMB ionisation constraints}

The cosmic microwave background (CMB) formed in the early stages of the universe at a redshift of $z\sim1100$ as the temperature dropped below the ionization energy of hydrogen. The CMB constitutes the surface of last scattering of the primordial photons, as at this moment the electrons and nucleons recombined to form neutral hydrogen atoms. The CMB has been free streaming to us from this moment and allows us to observe a snapshot of the universe as it was a mere 380 000 years old. However, since our DM candidate can annihilate into charged SM particles, it can reintroduce a certain amount of ionization during and after recombination. This amount of extra ionization can be constrained with the CMB since it would distort the theoretically clean CMB polarization spectrum.

\begin{figure}[!t]
\begin{centering}
    \includegraphics[width=0.49\textwidth]{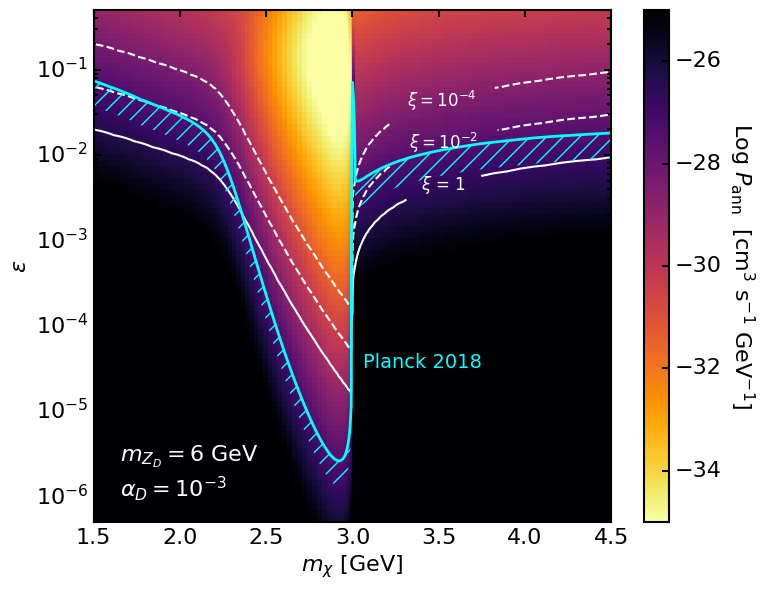} %
    \includegraphics[width=0.49\textwidth]{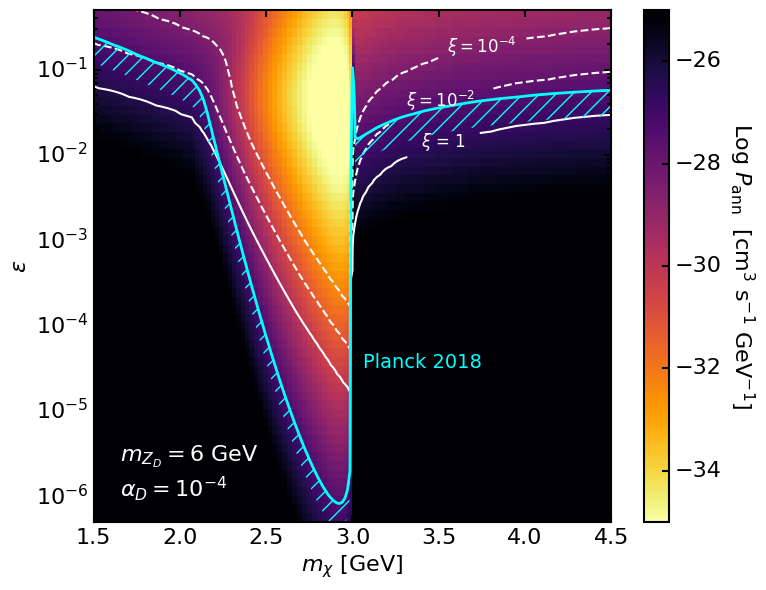} \\
    \includegraphics[width=0.49\textwidth]{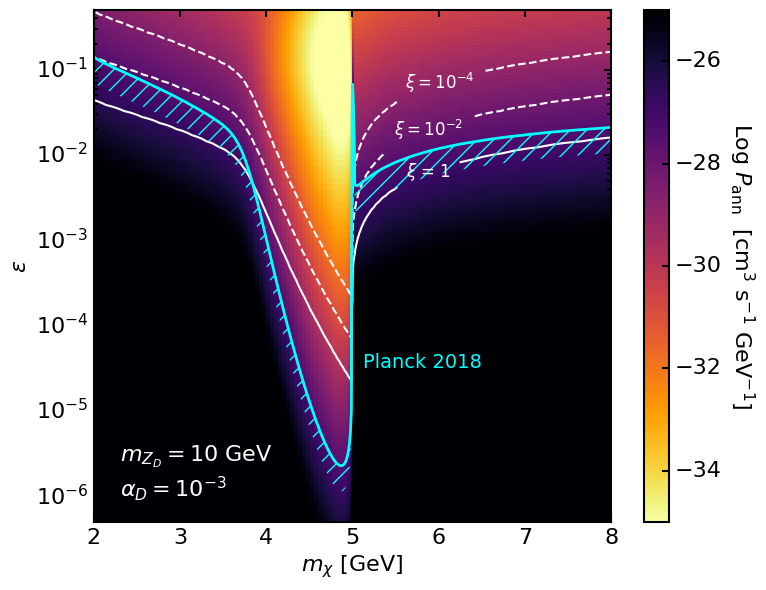} %
    \includegraphics[width=0.49\textwidth]{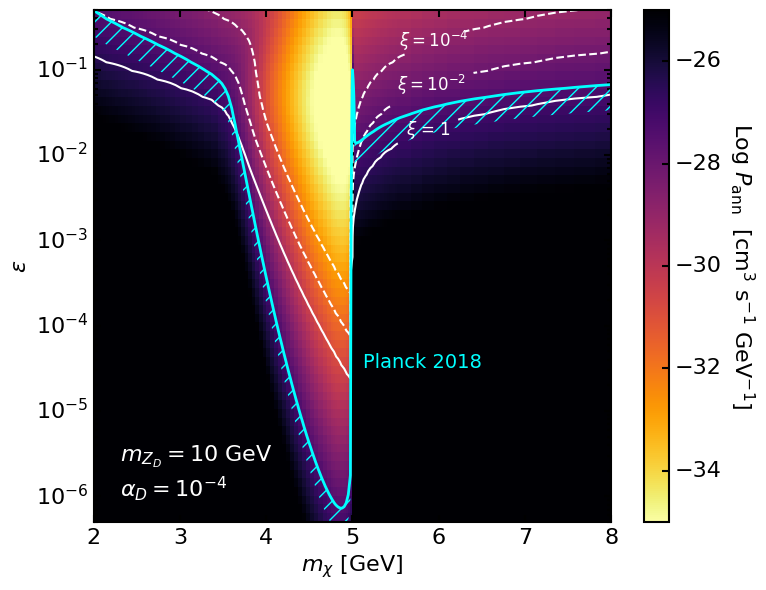}
    \caption{The colour map shows the values of DM annihilation parameter $P_\mathrm{ann}$ for dark photon masses of $m_{Z_D} = 6$~GeV (top) and $m_{Z_D} = 10$~GeV (bottom)  and couplings of $\alpha_D=10^{-3}$ (left) and $\alpha_D=10^{-4}$ (right). The cyan line denotes the limits on DM annihilation from Planck measurements of CMB anisotropies quoted in~\cref{eq:pcmb}. Note that all region of parameter space \textit{below}  this limit (i.e.~smaller $\epsilon$) is excluded.}
    \label{fig:cmb}
\end{centering}
\end{figure}

The rate of released energy into the primordial bath of particles from DM annihilations  that proceeds into ionization can be quantified as~\cite{Finkbeiner:2011dx}
\begin{equation}\label{eq:ion}
    \frac{d^2E}{dV \, dt} =  \rho_{\chi}^2\, (1+z)^6\ P_\mathrm{ann}(z) \equiv \xi^2\, \rho_\mathrm{CDM}^2\, (1+z)^6\ \sum_i f_i(z) \ \frac{\langle\sigma v\rangle_i}{m_\chi}\,,
\end{equation}
where the model dependence of the DM annihilation is usually quantified by the annihilation parameter
\begin{equation}
    P_\mathrm{ann}(z) = \sum_i f_i(z) \ \frac{\langle\sigma v\rangle_i}{m_\chi} \,.
\end{equation}
Here, $\rho_{\chi}$ is the DM density stored in $\chi$ particles at the formation of the CMB, $m_\chi$ the DM mass, and $\langle\sigma v\rangle_i$  and $f(z)_i$ denote the thermal average annihilation cross-section and redshift-dependent ionisation efficiency factor for the primary annihilation channel $i$. These efficiency factors for ionisation through low-energy  photons and $e^+e^-$ pairs  have been accurately computed in~\cite{Slatyer:2015kla,Liu:2016cnk}. As it turns out, the energy injection into the primordial gas is most efficient at around a redshift of $z\sim600$~\cite{Galli:2011rz,Slatyer:2015jla}. So we can compute some approximate mass-dependent efficiency factors for all the primary annihilation channels of $\chi$ as
\begin{align}
    f_i(m_\chi) = \int_0^{m_\chi} \frac{E\, dE}{2m_\chi} \left[2\, f^{e^+e^-}_\mathrm{eff}(E)\ \left(\frac{dN_{e^+}}{dE}\right)_i + f^{\gamma}_\mathrm{eff}(E)\ \left(\frac{dN_{\gamma}}{dE}\right)_i\right] \,,
\end{align}
where $f^c_\mathrm{eff}(E) \equiv f^c(E, z=600)$ denote the efficiency factors at a redshift of $z=600$ and $(dN_c/dE)_i$ are the secondary photon and positron spectra  produced from EM showers and decays of the primary particles $i$. For these spectra we rely on the tabulated values in \href{http://www.marcocirelli.net/PPPC4DMID.html}{\texttt{PPPC4DMID}} provided in Ref.~\cite{Cirelli:2010xx}.

Finally, taking into account the scaling of~\cref{eq:ion} with the dilution factor, we can define an effective, mass-dependent annihilation parameter as
\begin{equation}
    P_\mathrm{eff}(m_\chi) = \xi^2\  \sum_i f_i(m_\chi) \ \frac{\langle\sigma v\rangle_i}{m_\chi}\,.
\end{equation}
This allows us to directly set bounds on the amount of extra ionisation injected by our DM candidate from the constraint set by the Planck mission on the annihilation parameter from CMB anisotropies of~\cite{Planck:2018vyg}
\begin{equation}\label{eq:pcmb}
    P_\mathrm{ann} < 3.5 \times 10^{-28} \ \mathrm{cm}^3\, \mathrm{s}^{-1}\ \mathrm{GeV}^{-1}\,.
\end{equation}

In~\cref{fig:cmb}, we show the resulting values of the annihilation parameter for the same four benchmark scenarios considered in the last section. The Planck exclusion limit is shown by the cyan curve, which rules out all the parameter space below this curve. For reference, we show the relic abundance contours for $\xi\in\{1, 10^{-2}, 10^{-4}\}$. Similarly to the gamma ray signal discussed in the last section, we can clearly observe the same $\epsilon$-scaling behaviour of the annihilation parameter $P_\mathrm{eff}$, which rapidly increases for decreasing $\epsilon$ (or, equivalently, increasing $\xi$).

\subsection{Collider probes of a GeV-scale (invisible) dark photon}
\label{sec:coll}

The \dmmodel has a rich collider phenomenology. In this work we concentrate on the signatures related to the dark photon, and defer a study of dark Higgs boson $h_D$ signatures for a forthcoming work (see~\cite{Duerr:2017uap,Ferber:2023iso} for works in this direction). We also focus on the parameter space region for which the dark photon invisible decay mode into DM final states  $Z_D \to \chi \bar{\chi}$ is kinematically open and dominates the decay width. In this case, existing limits from dark photon searches in visible final states (e.g.~di-lepton searches by BaBar~\cite{BaBar:2014zli}, LHCb~\cite{LHCb:2019vmc} and CMS~\cite{CMS:2019buh,CMS:2023hwl}) get significantly weakened,\footnote{Besides the reduction of visible branching fractions, the presence of the invisible decay mode also widens the dark-photon resonance, which further weakens the limits from resonant searches of dark photons in visible final states~\cite{Felix:2025afw}.} and instead mono-$X$ signatures become an important probe of such invisible dark photon, provided that $\alpha_D \gg  \alpha_{\rm EM}\, \epsilon^2$.  For $m_{Z_D} < 10$~GeV, mono-photon searches at BaBar~\cite{BaBar:2017tiz} constrain the value of the kinetic mixing parameter to $\epsilon \lesssim 10^{-3}$ at the 90\% \cl. However, above the reach of $B$-factories the existing collider (LEP and LHC) bounds on $\epsilon$ are much less stringent, and we discuss them in the following.

\begin{figure}[!t]
    \begin{centering}
    \includegraphics[width=0.27\textwidth]{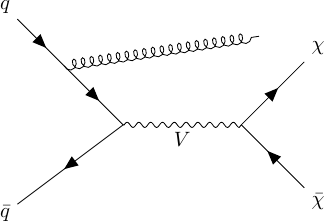} 
    \hspace{5mm}
    \includegraphics[width=0.27\textwidth]{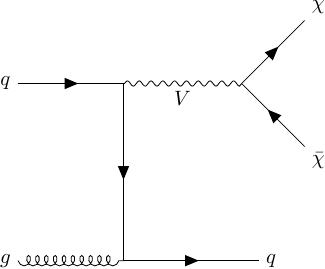} 
    \hspace{5mm}
    \includegraphics[width=0.27\textwidth]{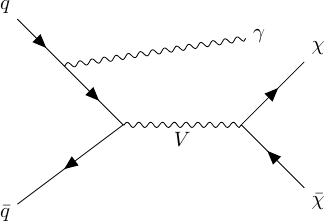} 
    \caption{Feynman diagrams contributing to mono-jet (left and centre) and mono-photon (right) in the dark Abelian Higgs portal to DM model, with $V = Z_D$, $Z$.}
    \label{fig_Feynman_monoX}
\end{centering}
\end{figure}

\subsubsection{LHC mono-jet and LEP single-photon searches}
\label{section:monojet}

Searches for events with (multi)-jets produced in association with a sizable amount of missing transverse momentum $p_{T}^{\rm miss}$, commonly referred to as {\it mono-jet} searches, may be the current leading probe of an invisibly decaying dark photon at the LHC. Mono-jet events are produced in our model via the process $p p \to \chi \bar{\chi} + j$, via $s$-channel $Z_D$ production and $s$-channel $Z$ production (see, e.g., the left and centre panel of~\cref{fig_Feynman_monoX}). The mono-jet production cross-section scales as $\propto \epsilon^2$. Very recently, the CMS Collaboration has published an interpretation of their mono-jet results~\cite{CMS:2021far} with $\sqrt{s} = 13$ TeV and $137$ fb$^{-1}$ of integrated luminosity for a minimal invisibly decaying dark photon model\footnote{The CMS mono-jet search~\cite{CMS:2021far} does include an interpretation for simplified spin-$1$ (either vector or axial-vector) mediator to fermionic DM models, closely related to our invisible dark photon model. Yet, re-using these limits in our scenario is not possible, since the dark photon inherits its coupling to SM fermions via mixing with the SM $Z$ boson, which has both vectorial and axial couplings to SM fermions. } (see section 7.1.1 of Ref.~\cite{CMS:2024zqs}) yielding 95\% C.L. exclusion limits on the kinetic mixing parameter as a function of $m_{Z_D}$.

Yet, such CMS dark-photon public limits are very coarse-grained, particularly for $m_{Z_D}$ in the vicinity of $m_{Z}$. To obtain a smooth limit from CMS mono-jet searches as a function of $m_{Z_D}$, we use a modified version of the minimal dark photon \texttt{UFO}~\cite{Degrande:2011ua} model developed in Refs.~\cite{Curtin:2013fra,Curtin:2014cca}, to which we add a Dirac dark fermion. We simulate $\sqrt{s} = 13$ TeV LHC $p p \to \chi \bar{\chi} + j$ parton-level mono-jet events in \texttt{MadGraph5\_aMC@NLO}~\cite{Alwall:2014hca}, with $p_{T}^{j} \equiv p_{T}^{\rm miss} > p_T^{\rm MIN}=  250$~GeV (this cut defines the signal region in the CMS mono-jet analysis~\cite{CMS:2021far}).

\begin{figure}[!t]
\begin{centering}
\includegraphics[width=0.465\textwidth]{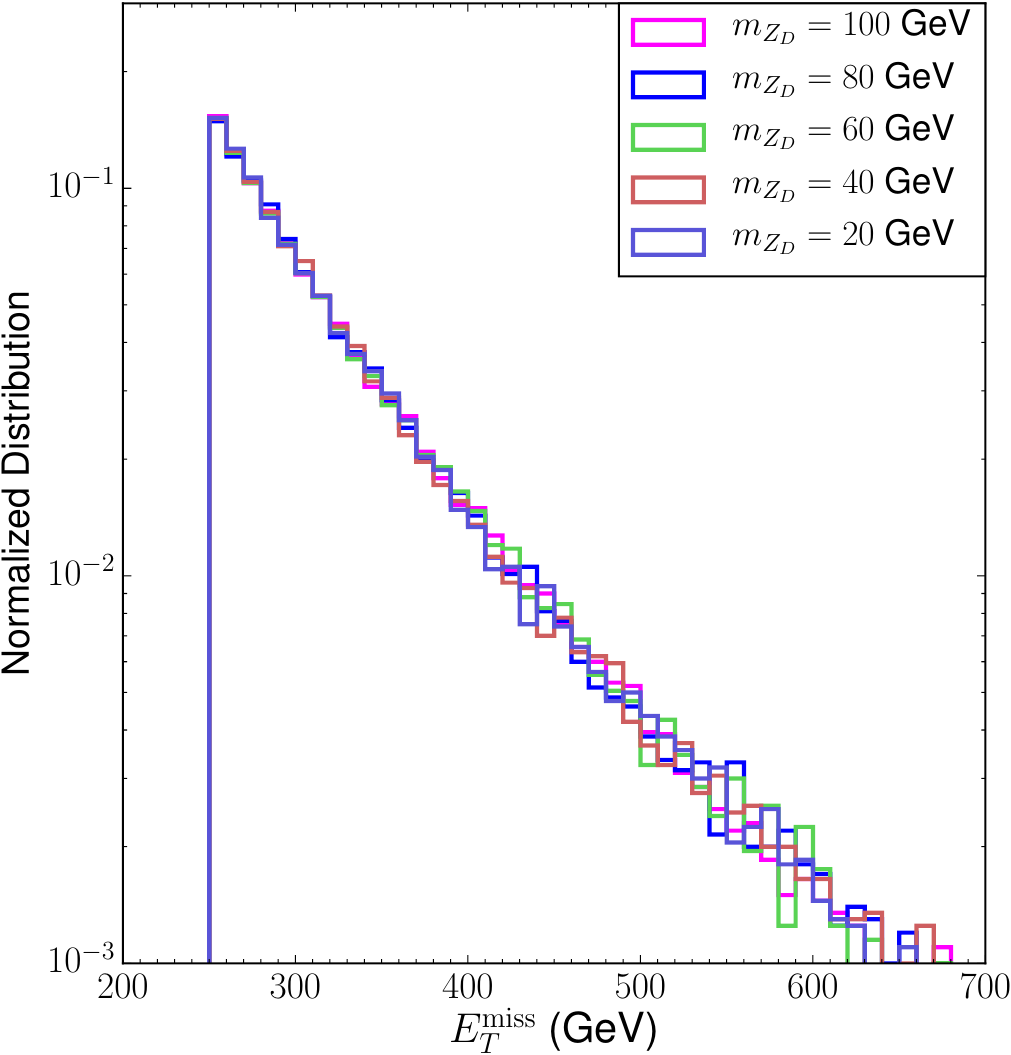}
    \raisebox{-3mm}{\includegraphics[width=0.5\textwidth]{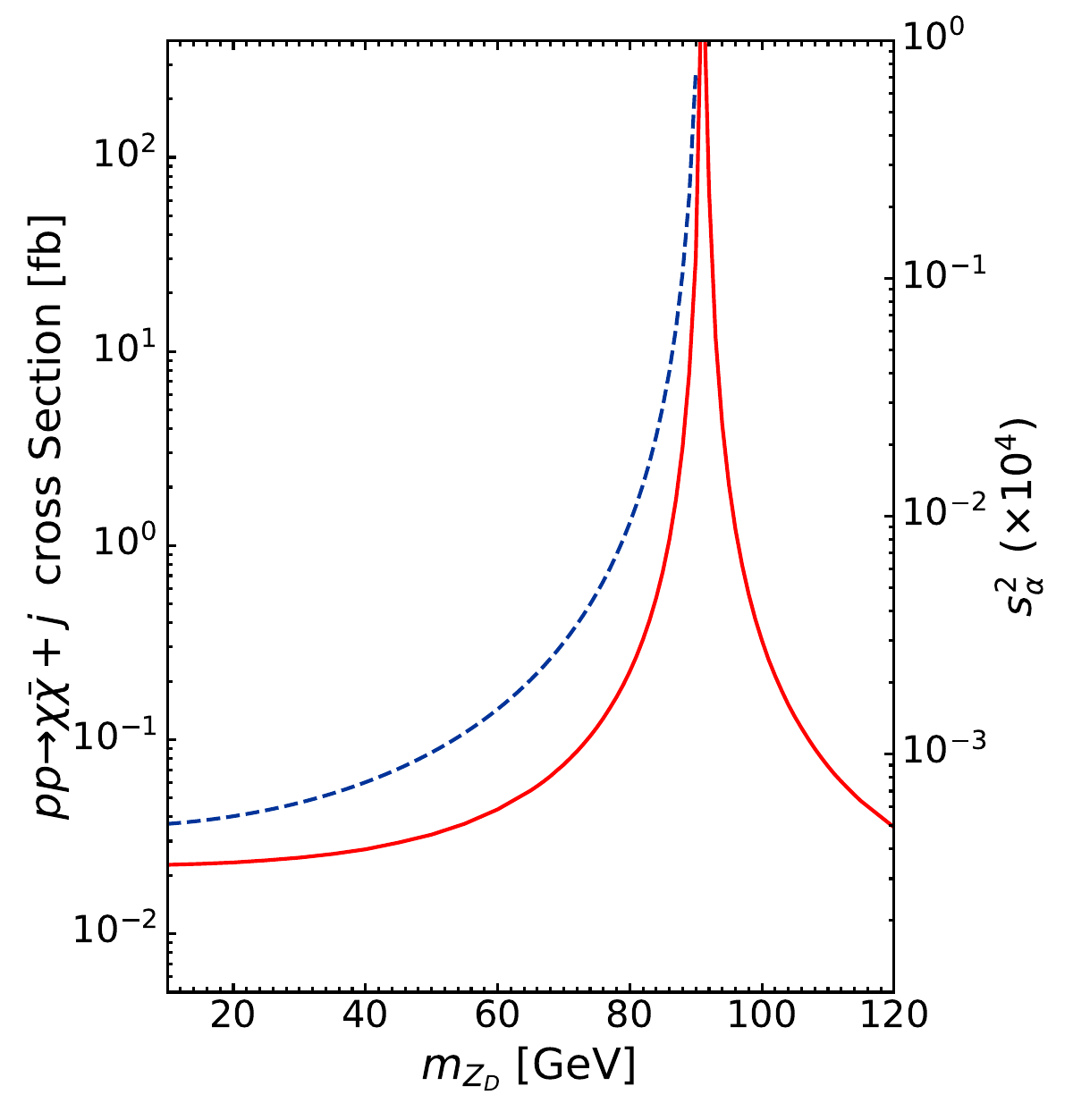}}
    \caption{Left: Parton level $p_{T}^{j} \equiv E_{T}^{\rm miss}$ normalized differential distributions for $p p \to \chi \bar{\chi} + j$ at $\sqrt{s} = 13$ TeV LHC, for different values of $m_{Z_D}$, with a cut $p_{T}^j > 250$~GeV. Right: Parton-level cross-section for $p p \to \chi \bar{\chi} + j$ at $\sqrt{s} = 13$ TeV LHC, for $\epsilon = 10^{-3}$ and $p_{T}^j > 250$~GeV (solid-red line), and squared dark photon mixing angle $s_{\alpha}^2$ (see \cref{gauge_mixing_angle}) as a function of $m_{Z_D}$ (for $m_{Z_D} < m_Z$) with fixed $\epsilon = 10^{-3}$ (dashed-blue line).}
    \label{fig1:XS_Monojet}
\end{centering}
\end{figure}

In~\cref{fig1:XS_Monojet}, we show the  normalized $p_{T}^{j}$ differential distribution for different values of $m_{Z_D}$ (left) and the mono-jet cross-section as a function of $m_{Z_D}$ for a reference value $\epsilon = 10^{-3}$  (right). For dark photon masses $m_{Z_D} \ll p_T^{\rm MIN}$, the specific value of $m_{Z_D}$ has very little impact on the $p_{T}^{j}$ kinematic distribution (as seen explicitly in the left panel of~\cref{fig1:XS_Monojet}) and the change in the experimental limit on the kinetic mixing $\epsilon$ is mostly dictated by the corresponding change in the mono-jet production cross-section as $m_{Z_D}$ varies (it becomes very large around $m_Z$ and then decreases to reach a fairly constant value for $m_{Z_D} \ll m_Z$, as shown in the right panel of~\cref{fig1:XS_Monojet}).
This behaviour of the cross-section is due to the dependence of the dark-photon mixing $s_{\alpha}$ in \cref{gauge_mixing_angle} on $m_{Z_D}$ (for fixed $\epsilon$), shown in the right panel of~\cref{fig1:XS_Monojet} as a dashed-blue line.

We then perform a cross-section rescaling of the mono-jet CMS limit on $\epsilon$ obtained for $m_{Z_D} \ll m_Z$ to the entire range $m_{Z_D} < m_Z$, and the corresponding smooth mono-jet current 95\% C.L. exclusion limit is shown on the left panel of \cref{fig:DarkPhoton_Collider}, which is in good agreement with the coarse-grained CMS public limits~\cite{CMS:2024zqs}. We also carry out a naive extrapolation of the sensitivity of mono-jet searches to High-Luminosity (HL) LHC, by simply accounting for the statistical increase in sensitivity from the present mono-jet search with $137$~fb$^{-1}$ to the HL-LHC with $3000$~fb$^{-1}$ of integrated luminosity. This would amount to a rescaling of the sensitivity to the $p p \to \chi \bar{\chi} + j$ signal cross-section by a factor of $\sqrt{137/3000}$. Taking into account that the signal cross-section scales as $\epsilon^2$, yields the naive HL-LHC 95\% C.L. mono-jet sensitivity projection shown on the left panel of \cref{fig:DarkPhoton_Collider}. We stress that both the current mono-jet limits and the HL-LHC projection assume a branching fraction of BR$(Z_D \to \chi \bar{\chi}) = 1$.

Turning now to single-photon searches at LEP~\cite{DELPHI:2003dlq,L3:2003yon,DELPHI:2008uka}, these also constrain scenarios of the dark Abelian Higgs portal coupled to DM for $m_{Z_D} > 2 m_{\chi}$. In this case, single-photon events are produced via diagrams like the one on the right in~\cref{fig_Feynman_monoX}. Assuming again that BR$(Z_D \to \chi \bar{\chi}) = 1$, the relevant limits on the kinetic mixing $\epsilon$ as a function of $m_{Z_D}$ can be conveniently obtained with DarkCast~\cite{Ilten:2018crw}, which uses an interpolation of the corresponding limits derived in Ref.~\cite{Fox:2011fx} using the LEP data from DELPHI~\cite{DELPHI:2003dlq,DELPHI:2008uka}. We show the resulting LEP single-photon limits on the left panel of \cref{fig:DarkPhoton_Collider}. It can be seen that these are stronger than current CMS mono-jet limits, while the sensitivity of mono-jet searches at the HL-LHC may yield comparable constraining power to LEP single-photon bounds (particularly if the cross-section sensitivity exceeds the naive $\sqrt{\mathcal{L}}$ statistical increase, see Ref.~\cite{Belvedere:2024wzg} for a discussion).

\subsubsection{Electroweak precision observables}
\label{section:EWPO}

\begin{figure}[!t]
\begin{centering}
    \includegraphics[width=0.99\textwidth]{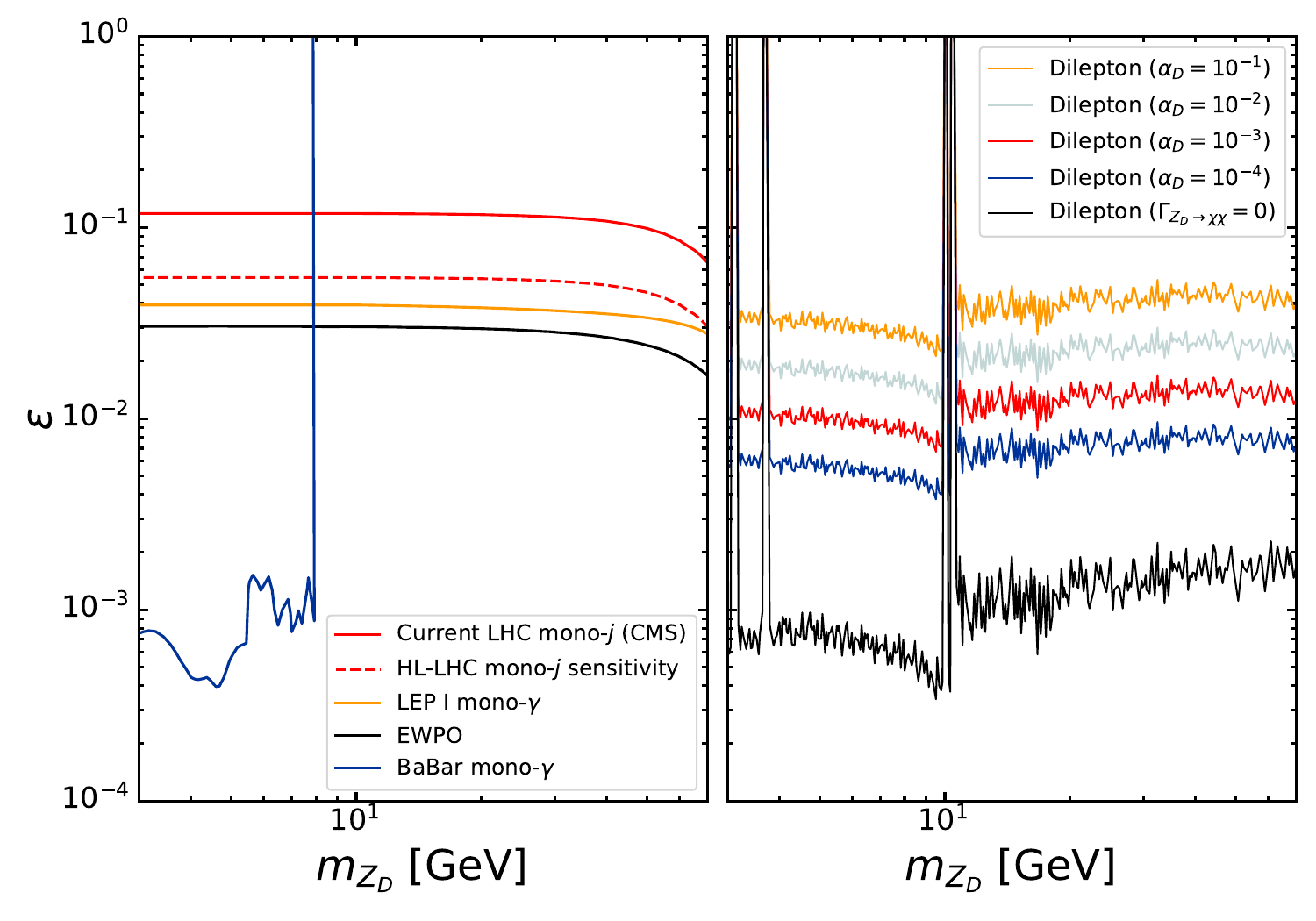} 
    \caption{Left: Current limits on $\epsilon$ as a function of $m_{Z_D}$ from EWPO (solid-black), from LEP mono-photon searches (solid-yellow), from BaBar mono-photon searches (solid-dark-blue) and from LHC mono-jet searches by CMS (solid-red).  Also shown is the naive $\sqrt{\mathcal{L}}$ extrapolation of the mono-jet CMS sensitivity to HL-LHC (dashed-red). All the direct searches assume BR$(Z_D \to \chi \bar{\chi}) = 1$. Right: 90\% C.L. upper limits on $\epsilon$ from di-lepton searches by BaBar, LHCb and CMS as a function of $m_{Z_D}$, respectively for $\Gamma_{Z_D\to \chi\chi} = 0$ (black), $\alpha_D = 10^{-1}$ (yellow), $\alpha_D = 10^{-2}$ (grey), $\alpha_D = 10^{-3}$ (red), $\alpha_D = 10^{-4}$ (dark-blue). All curves (except the $\Gamma_{Z_D\to \chi\chi} = 0$ one) are computed for $m_{\chi} = m_{Z_D}/3$.
    }
    \label{fig:DarkPhoton_Collider}
\end{centering}
\end{figure}

In addition to direct searches for an invisibly-decaying $Z_D$ at LEP and the LHC, electroweak precision observables (EWPO) are a sensitive probe of the mixing between the SM $Z$-boson and the dark photon, yielding an indirect constraint on the kinetic mixing parameter. This has been studied thoroughly in the literature (see, e.g., Refs.~\cite{Babu:1997st,Hook:2010tw,Curtin:2014cca,Bernreuther:2022jlj,Cheng:2022aau,Zeng:2022lkk,Bento:2023weq}). In this work, we do not attempt a global fit of EWPO in the presence of kinetic mixing (such as the one of Ref.~\cite{Curtin:2014cca}), but rather focus on the effect of $\epsilon$ on the oblique EW corrections $\alpha_{\rm EM} S$ and $\alpha_{\rm EM} T$, following the analysis of Ref.~\cite{Babu:1997st}. 
We start by writing the neutral current interactions of the physical $Z$-boson with a SM fermion $f_i$ as~\cite{Peskin:1991sw,Burgess:1993vc}
\begin{align}
\label{eq:lag_oblique}
    \mathcal{L}_{\rm NC}  =\frac{-e}{s_W c_W}\left(1+\frac{\alpha_{\rm EM} T}{2}\right) \overline{f_i} \gamma^\mu \left[T_{3_i} P_L-Q_i\left(s_w^2+\frac{\alpha_{\rm EM} S}{4\left(c_W^2-s_W^2\right)}-\frac{c_W^2 s_W^2 \alpha_{\rm EM} T}{c_W^2-s_W^2}\right)\right] f_i  \ Z_\mu \, ,
\end{align}
with $Q_i$ the electric charge of the corresponding SM fermion.
We can then match the gauge-fermion interactions in the \dmmodel (in the limit $g_D \to 0$) with~\cref{eq:lag_oblique} to extract $\alpha_{\rm EM} S$ and $\alpha_{\rm EM} T$. From~\cref{eq:fermion_currents} we get (see also Ref.~\cite{Chun:2010ve}) 
\begin{align}
\label{eq:NC}
    \mathcal{L}_{\rm NC}  = \frac{-e}{s_W c_W}\, c_{\alpha} \, \overline{f_i} \gamma^\mu\left[T_{3_i} P_L  \left(1 -  \frac{s_{\alpha}}{c_{\alpha}} \delta \right)  -Q_i\left(s_W^2  - \frac{s_{\alpha}}{c_{\alpha}} \delta  \right)\right] f_i \, Z_\mu \, ,
\end{align}
from which we can readily obtain $S(\epsilon, m_{Z_D})$ and $T(\epsilon, m_{Z_D})$ by comparing to \cref{eq:lag_oblique}. We then use the best-fit values and standard deviations from Ref.~\cite{ParticleDataGroup:2024cfk} for $S$ and $T$ (assuming $U = 0$), 
\begin{align}
\label{chi_EWPO1}
    S  &= -0.05 \pm 0.07\, ,  \nonumber \\
    T  &= 0.00 \pm 0.06\, ,
\end{align}
to derive the bounds from EWPO. We perform a $\chi^2$ fit to $\mathcal{O}_{1} = S(\epsilon, m_{Z_D})$ and $\mathcal{O}_{2} = T(\epsilon, m_{Z_D})$, given by 
\begin{equation}
\label{chi_EWPO2}
    \Delta\chi^2 (\epsilon,m_{Z_D}) = \sum_{i,j} \left(\mathcal{O}_{i}(\epsilon,m_{Z_D}) - \mathcal{O}^0_{i}\right) (\sigma^2)_{ij}^{-1}
    \left(\mathcal{O}_{j}(\epsilon,m_{Z_D}) - \mathcal{O}^0_{j}\right)\, ,
\end{equation}
with $\mathcal{O}^0_{i}$ denoting the central values in \cref{chi_EWPO1} and $(\sigma^2)_{ij} \equiv \sigma_i \rho_{ij} \sigma_j$, being $\sigma_i$ the respective standard deviations from \cref{chi_EWPO1}. The covariance matrix, $\rho_{ij}$, in the $S-T$ plane is given by~\cite{Haller:2018nnx}
\begin{equation}
\rho_{ij} = \left(\begin{array}{cc}
             1 & 0.92\\
             0.92 &1
            \end{array}\right)\, .
\end{equation}
Our resulting 95\% C.L. EWPO limits on $\epsilon$ as a function of $m_{Z_D}$ (given by $\Delta\chi^2 (\epsilon,m_{Z_D}) = 4$) agree well with those obtained in Ref.~\cite{Curtin:2014cca}, and are shown in the left panel of \cref{fig:DarkPhoton_Collider} as a solid black-line. They yield the strongest current limits for a fully invisible dark photon (with BR$(Z_D \to \chi \bar{\chi}) = 1$) above the reach of $B$-meson factories like BaBar.

\subsubsection{Visible (di-lepton) searches}
\label{section:dilepton}

The constraints on an invisible dark photon discussed in section~\ref{section:monojet} above assume {BR$(Z_D \to \chi \bar{\chi}) \simeq 1$}, corresponding to the $\alpha_D \gg \alpha_{\rm EM} \epsilon^2$ region of the DAHM-DM parameter space. In this region, the existing constraints on the kinetic mixing from di-lepton final states from BaBar~\cite{BaBar:2014zli},  LHCb~\cite{LHCb:2019vmc} and CMS~\cite{CMS:2019buh,CMS:2023hwl} become weaker due to the presence of the dominant invisible contribution $\Gamma_{Z_D\to \chi\chi}$ to the dark photon decay width $\Gamma_{Z_D} = \Gamma_{Z_D\to {\rm SM}} +  \Gamma_{Z_D\to\chi\chi}$. Yet, given the strong sensitivity of di-lepton searches, they can still be important even when diluted by an invisible partial width $\Gamma_{Z_D\to \chi\chi}$, in particular if $\alpha_D \gg \alpha_{\rm EM} \epsilon^2$ does not hold. 
We compute here the modified limits from di-lepton searches, which are simply given by
\begin{equation}
\label{eq:weak_epsilon_dilepton}
    \epsilon < \epsilon_{90}^{\ell\ell} \times \sqrt{\frac{\Gamma_{Z_D\to {\rm SM}} +  \Gamma_{Z_D\to\chi\chi}}{\Gamma_{Z_D\to {\rm SM}}}} ,
\end{equation}
where $\epsilon_{90}^{\ell\ell}$ is the existing 90\% C.L. di-lepton limit from the combination of BaBar, LHCb and CMS searches~\cite{BaBar:2014zli,LHCb:2019vmc,CMS:2019buh,CMS:2023hwl} for a dark photon with $\Gamma_{Z_D\to\chi\chi} = 0$. The partial width $\Gamma_{Z_D\to\chi\chi}$ is given by (see, e.g., Ref.~\cite{Fabbrichesi:2020wbt})
\begin{equation}
    \Gamma_{Z_D\to\chi\chi} = \frac{\alpha_D\, m_{Z_D}}{3} \sqrt{1- \frac{4 m_{\chi}^2}{m_{Z_D}^2}} \left[ 1 + 2\frac{m_{\chi}^2}{m_{Z_D}^2} \right]\, .
\end{equation}
For a GeV-scale dark photon $Z_D$, the partial width into Standard Model final states $\Gamma_{Z_D\to {\rm SM}}$ can be obtained with high accuracy following Refs.~\cite{Curtin:2014cca,Bauer:2018onh}. Considering the leptonic partial width~\cite{Fabbrichesi:2020wbt}   
\begin{equation}
    \Gamma (Z_D \to \ell \ell) \simeq \frac{\alpha_{\rm EM}\, \epsilon^2 \, m_{Z_D}}{3} \sum_{\ell = e,\, \mu} \sqrt{1- \frac{4 m_{\ell}^2}{m_{Z_D}^2}} \left[ 1 + 2\frac{m_{\ell}^2}{m_{Z_D}^2} \right] \, ,
\end{equation}
the value of $\Gamma_{Z_D\to {\rm SM}}$ may be obtained as 
\begin{equation}
\Gamma_{Z_D\to {\rm SM}} = \frac{\Gamma (Z_D \to \ell \ell)}{{\rm BR}^{(0)}_{Z_D \to \ell\ell}}  \, ,
\end{equation}
where BR$^{(0)}_{Z_D \to \ell\ell}$ is the leptonic branching fraction for a dark photon with $\Gamma_{Z_D\to \chi\chi} = 0$ for the range $m_{Z_D} \in [0.1,\, 100]$~GeV (shown explicitly in Fig.~2 (left) of Ref.~\cite{Curtin:2014cca}), obtained using $R \equiv \sigma(e^+ e^- \to {\rm hadrons})/\sigma(e^+ e^- \to \mu^+ \mu^-)$ experimental data~\cite{ParticleDataGroup:2024cfk}.

On the right panel of \cref{fig:DarkPhoton_Collider}, we show the di-lepton limits on $\epsilon$ for $\Gamma_{Z_D\to \chi\chi} = 0$, together with the corresponding limits derived from \cref{eq:weak_epsilon_dilepton} for $\alpha_D = 10^{-1}$, $10^{-2}$, $10^{-3}$, $10^{-4}$. The narrow gaps in the coverage of di-lepton searches depicted on the right panel of \cref{fig:DarkPhoton_Collider} correspond to $\pm 50$ MeV vetoes by the BaBar analysis~\cite{BaBar:2014zli} around the $J/\psi$, $\psi(2S)$, $\Upsilon(1S)$ and $\Upsilon(2S)$ resonances.\footnote{The LHCb di-lepton analysis~\cite{LHCb:2019vmc} imposes wider mass vetoes around the same resonances.} We see explicitly that the di-lepton limits, even when diluted, can surpass those from searches for invisible dark photons (mono-$X$), depicted on the left panel of \cref{fig:DarkPhoton_Collider}, for $\alpha_D \ll 0.1$, which is our main parameter region of interest in this work (recall the discussion in \cref{sec:dd,sec:id}). This is particularly the case as $m_\chi \to m_{Z_D}$, which can further suppress the $\Gamma_{Z_D\to \chi\chi}$ partial width (an effect not shown in \cref{fig:DarkPhoton_Collider}).

\section{Results}
\label{sec:res}

Having discussed all the relevant observational constraints from direct and indirect detection, as well as collider experiments, we now combine them in this section.

\begin{figure}[!t]
\begin{centering}
    \includegraphics[width=0.49\textwidth]{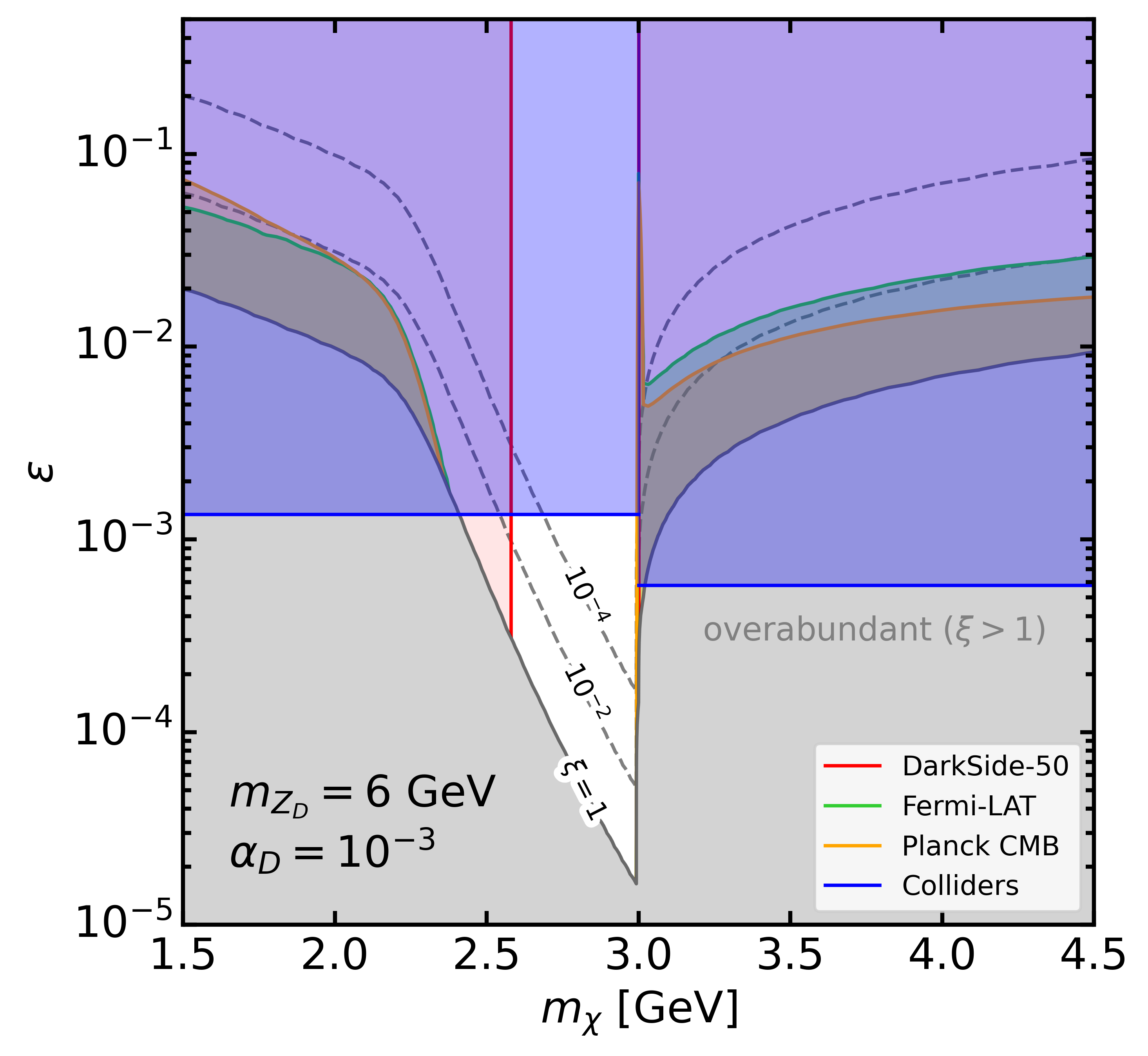} %
    \includegraphics[width=0.49\textwidth]{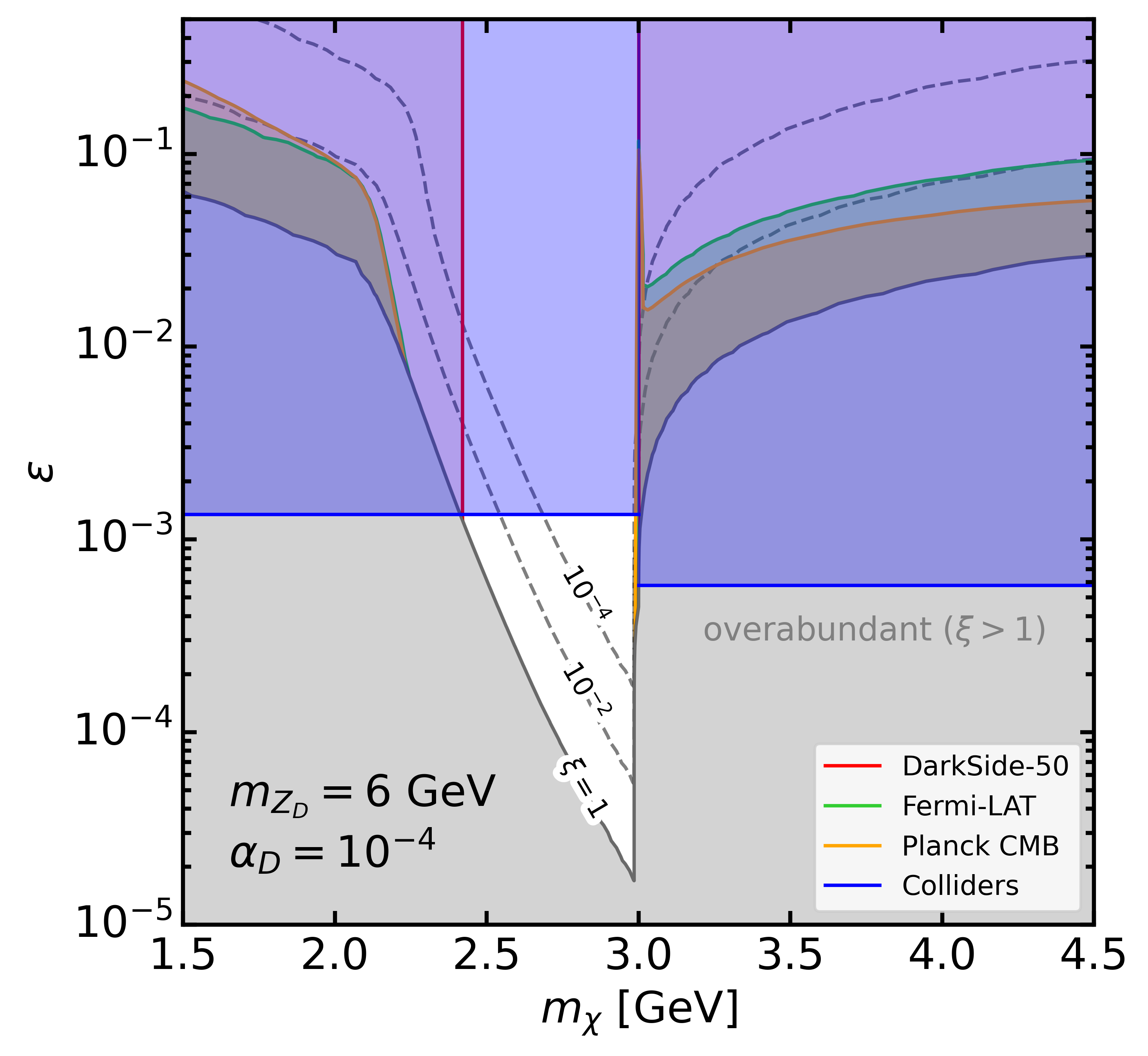} \\
    \includegraphics[width=0.49\textwidth]{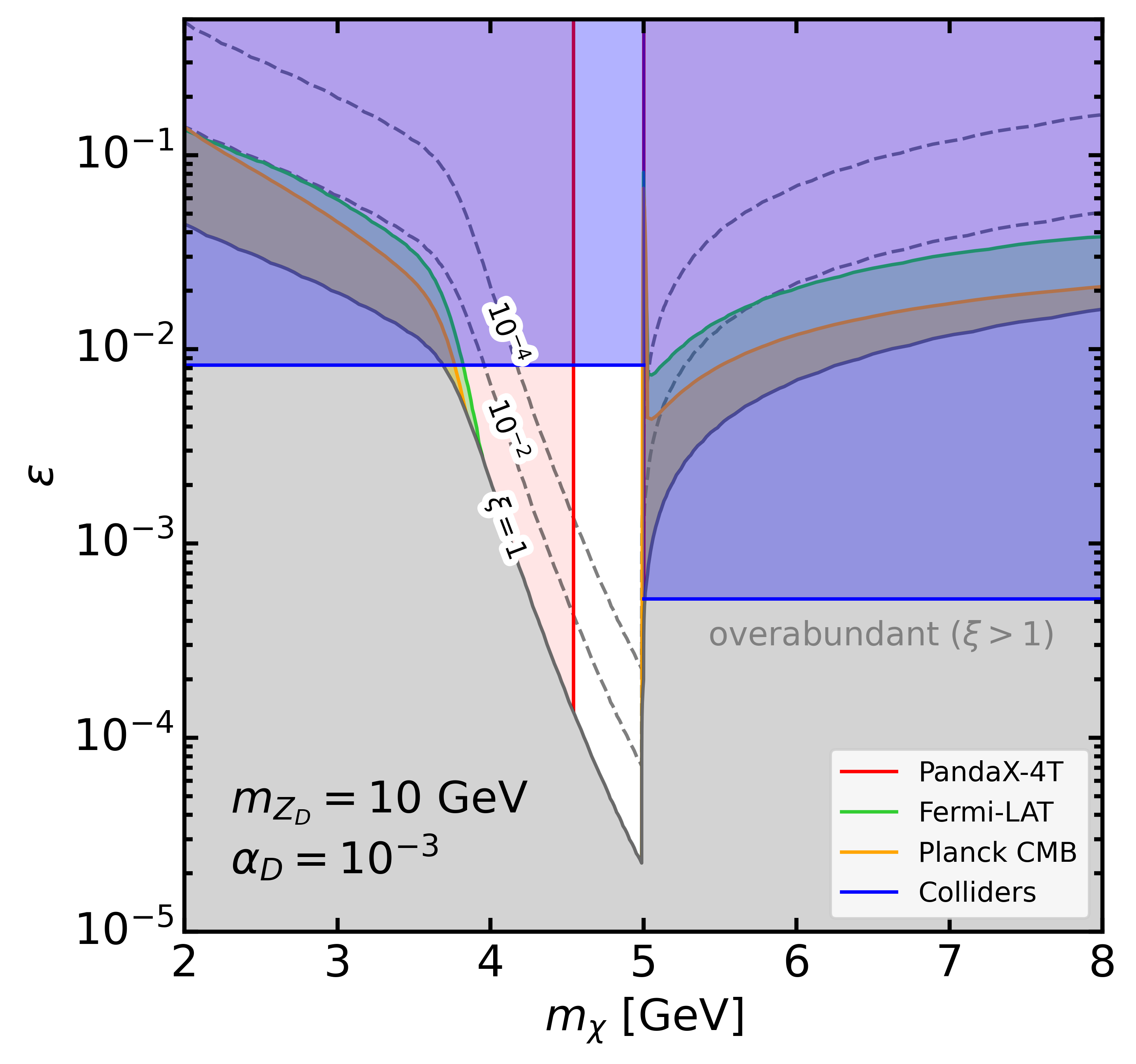} %
    \includegraphics[width=0.49\textwidth]{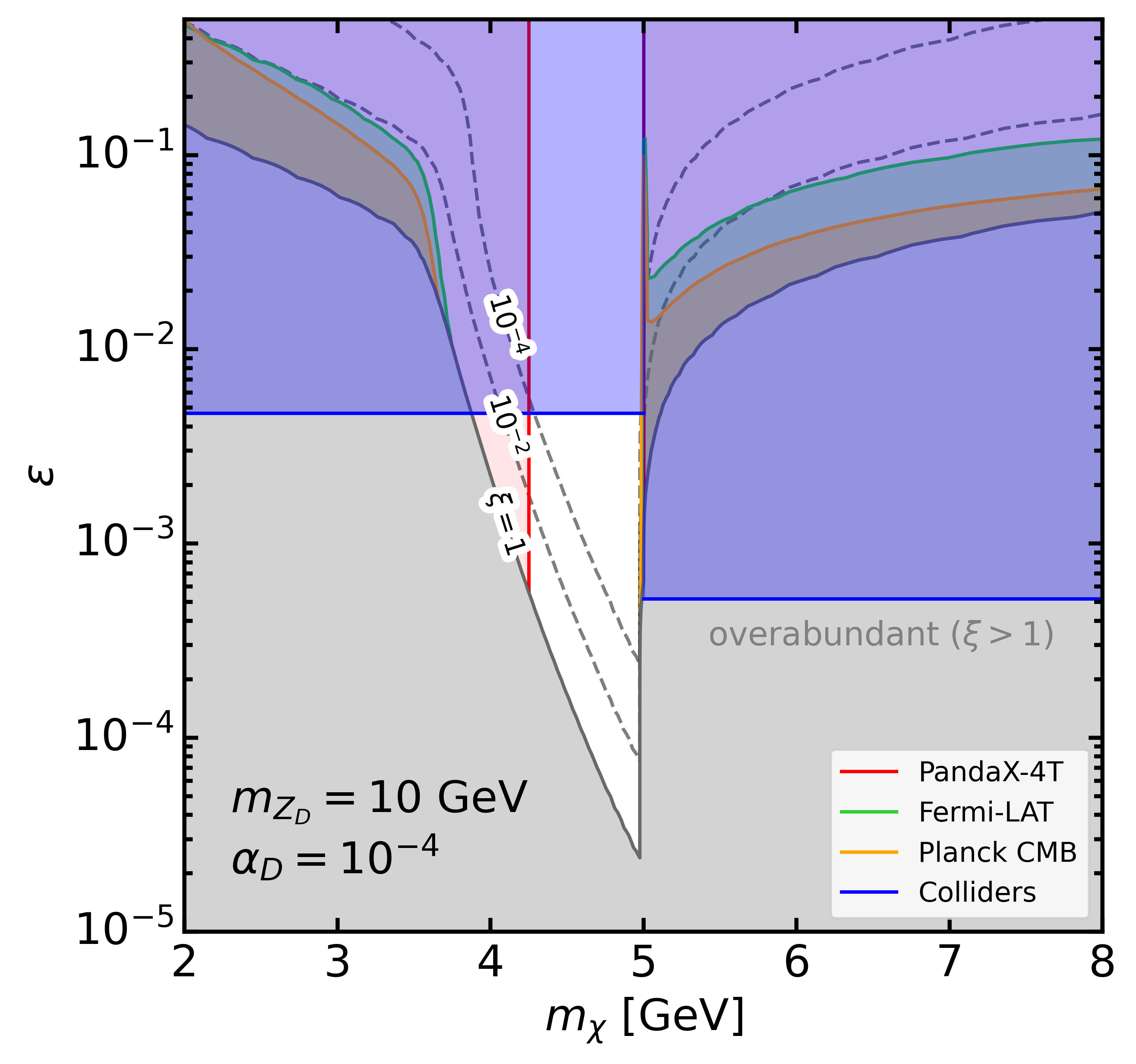}
    \caption{
    The white area represents the viable parameter space in the $(\epsilon,\,m_\chi)$ plane for vector-like fermion DM coupled to an invisible Dark Photon for a benchmark mass of $m_{Z_D}=6$  GeV (top) and $m_{Z_D}=10$  GeV (bottom). We show the constraints and the surviving parameter space for two different choices of the dark coupling constant, $\alpha_D=10^{-3}$ (left) and $\alpha_D=10^{-4}$ (right). The grey area is excluded because the relic abundance exceeds the Planck result. The red, green, yellow, and blue areas correspond to the areas excluded by direct detection, Fermi-LAT data on dSph, Planck data on CMB, and collider searches, respectively. More details are given in the main text.}
    \label{fig:results}
\end{centering}
\end{figure}

In~\cref{fig:results}, we show the resulting bounds in the $(\epsilon,\,m_\chi)$ plane for four different benchmark scenarios, with $m_{Z_D}=6\,(10)$~GeV in the top (bottom) row and $\alpha_D=10^{-3}\,(10^{-4})$ in the left (right) column, respectively.
The dashed grey lines represent different values of the relic abundance of $\chi$, and are labelled by the corresponding dilution parameter, $\xi = \{1, 10^{-2}, 10^{-4}\}$. Along the line with $\xi=1$, $\chi$ constitutes all of the observed DM in the universe, while in regions with $\xi<1$  it is only a subcomponent. Below the $\xi=1$ line (i.e.~where $\xi>1$), the gray area represents the region of the parameter space where the abundance of $\chi$ exceeds the amount of cold dark matter compatible with Planck data~\cite{Planck:2018vyg}, and is excluded for that reason.
The resonance peak in the thermal production cross-section at $m_\chi\sim m_{Z_D}/2$ is clearly visible as the resulting dip in the relic density lines.\footnote{It is worth noting that the asymmetric shape of the dip is due to thermal broadening of the Breit-Wigner resonance of the mediator. Due to the non-negligible  kinetic energies of the DM during thermal freeze-out in the early universe, the production cross-section can go on resonance even for DM candidates with $m_\chi < m_{Z_D}/2$~\cite{Griest:1990kh,Ibe:2008ye,Guo:2009aj}. On the other hand, for $m_\chi>m_{Z_D}/2$ the resonance can never exactly be hit, which leads to the strong sudden increase in relic density above the resonance.}

The most severe constraints arise from DD experiments, and are represented by the red shaded areas. As we saw in~\cref{sec:dd}, the allowed regions are narrow DM mass bands in the resonance dip, $m_\chi \sim m_{Z_D}/2$, which become broader when $\alpha_D$ decreases. As we explained in~\cref{sec:dd}, these bands are unconstrained by direct detection for any value of the kinetic mixing. For $m_{Z_D}=6$~GeV (upper panels in~\cref{fig:results}), the most stringent bound is due to the low-mass search by DarkSide-50~\cite{DarkSide-50:2022qzh}. For $m_{Z_D}=10$~GeV (lower panels in~\cref{fig:results}), the most stringent bound is due to the searches by PandaX-4T~\cite{PandaX-4T:2021bab}. This becomes clear by looking at the corresponding relic density lines shown together with the direct detection bounds in~\cref{fig:DDandRD}.

Indirect searches are less relevant in the resonance region. The green and yellow surfaces in~\cref{fig:results} represent the limits derived in~\cref{sec:id} from gamma ray searches in dSphs with Fermi-LAT and searches for energy injections into the CMB with Planck, respectively. As we explained in \cref{sec:id}, to derive these limits it is crucial to take into account the scaling of the relic abundance with the dilution factor $\xi$, which suppresses the indirect signals with increasing couplings. This leads to the relaxation of the Fermi-LAT and CMB constraints for underabundant DM (i.e.~larger couplings above the $\xi=1$ line).

Finally, as discussed in~\cref{sec:coll}, stringent bounds on invisibly decaying dark photons can be imposed from searches for missing energy in conjunction with a single photon at LEP~\cite{DELPHI:2003dlq,L3:2003yon,DELPHI:2008uka} and especially BaBar~\cite{BaBar:2017tiz} (for masses of $m_{Z_D}\lesssim 8$~GeV). Yet, the most sensitive probe of invisible dark photons in the mass range 10 GeV $\lesssim m_{Z_D}\lesssim m_{Z}$ is presently given by EWPO (c.f.~\cref{section:EWPO}). In addition, even when diluted by the invisible decay width, 
searches for di-lepton signatures at BaBar, LHCb and CMS can lead to more severe constraints on $\epsilon$ in the mass range of $m_{Z_D}\gtrsim10$~GeV for dark gauge couplings of $\alpha_D\lesssim10^{-2}$ (c.f.~\cref{fig:DarkPhoton_Collider}). In~\cref{fig:results}, we show these limits by means of the blue areas. Above the resonance, $m_{\chi}> m_{Z_D}/2$, the leading bound is due to the strong constraints on visible decays into di-leptons. In the invisible regime, $m_{\chi}\lesssim m_{Z_D}/2$, for $m_{Z_D}=6$~GeV the leading bound is given by the BaBar mono-photon search, while for $m_{Z_D}=10$~GeV still the visible channel di-lepton constraint dominates. While the exact value $m_{Z_D}=10$~GeV coincides with the BaBar veto region around the $\Upsilon(2S)$ resonance~\cite{BaBar:2014zli} (and thus the limit from di-lepton searches is not strictly applicable), a deviation of $\mathcal{O}(0.1)$~GeV from this exact value would make the di-lepton bounds applicable. We choose here to consider for the lower panels of~\cref{fig:results} the di-lepton bounds on the lower edge of the $\Upsilon(2S)$ veto region. As we can see, these collider searches provide a  complementary search  strategy to direct and indirect detection searches, boxing in the remaining allowed parameter space from above and constraining the remaining parameter space for sub-component  DM ($\xi<1$).

Combining all the bounds leaves only small, well-constrained islands in parameter space, (the remaining white stripes in~\cref{fig:results}) close to the resonance. In these regions, the vector-like fermion $\chi$ can constitute all of the DM, but also be a subdominant component (with decreasing density as the kinetic mixing increases).

Finally, note that, in the vicinity of the resonance, the standard freeze-out treatment~\cite{Steigman:2012nb} based on the number density Boltzmann equation and the assumption of kinetic equilibrium may receive sizable corrections, as discussed in Refs.~\cite{Binder:2017rgn,Binder:2021bmg}. In this regime, departures from kinetic equilibrium can modify the predicted relic abundance by up to one order of magnitude exactly at the resonance or close to it and mainly for specific choices of parameters, in particular for heavy final-state fermions~\cite{Binder:2021bmg}. In our scenario, where several fermionic channels contribute and the dominant annihilation modes involve light SM fermions, such effects are expected to be less pronounced. Moreover, even in the most conservative case, the relic density would be modified by less than one order of magnitude, corresponding to a shift of less than a factor $\sqrt{10}\approx3.16$ in the preferred value of the kinetic mixing. This would slightly narrow the allowed regions close to the resonance, but would not qualitatively affect the existence of viable parameter space.

\section{Conclusions}
\label{sec:conclusions}

In this work, we have considered a minimally UV-complete setup of a fermionic DM candidate, $\chi$, charged under a dark Abelian \Ux gauge symmetry that is spontaneously broken by the VEV of a dark scalar $S$. We have focused on mediator masses in the range of a few GeV up to the $Z$ mass, and we have derived observational constraints from direct and indirect DM searches, as well as collider limits. When comparing with DM observables, we have taken into account in a consistent way the possible DM density dilution in scenarios where $\chi$ does not account for all the DM in the universe. Our main results are shown in~\cref{fig:DDandRD,fig:DDandRD_HighMass,fig:results}, and we summarise them here.

\begin{itemize}
    \item In agreement with previous results in the literature, we have found that scenarios with large dark sector couplings, $\alpha_D\gtrsim 10^{-2}$, are ruled out, mainly by DD constraints, even in the resonance regime $m_\chi\sim m_{Z_D}/2$. However, for moderately smaller dark couplings, $10^{-2}\gtrsim\alpha_D\gtrsim 10^{-5}$ (or equivalently $10^{-1}\gtrsim g_D \gtrsim 10^{-2}$), we have found that there are viable regions of parameter space close to the resonance, $m_\chi\sim m_{Z_D}/2$, where $\chi$ can constitute all of the DM or be a subcomponent.

\item Indirect detection constraints are only important for values of the dilution factor $\xi$ close to one (for which $\chi$ makes up all the dark matter), and away from the resonant annihilation region. When the relic density of $\chi$ decreases, their $\xi^2$ scaling substantially weakens these constraints, allowing for large values of the kinetic mixing $\epsilon$, at the expense of having $\chi$ as a subdominant DM component.

\item In the future, next-generation DD experiments based on liquid noble gases such as DarkSide-LowMass or DARWIN will be able to probe most of the remaining allowed resonant DM  scenarios, improving the constraints by about two orders of magnitude in the dark gauge coupling, to $\alpha_D \lesssim 10^{-4}$. Below this value, any resulting DM signal would already be hidden in the neutrino fog, and a potential detection will become significantly more challenging. These searches will therefore narrow down the allowed parameter space in the $m_{Z_D}$ direction, while being insensitive to the value of the kinetic mixing.

\item Current and future collider searches for missing energy signals can probe complementary regions of the remaining parameter space with sizeable kinetic mixing $\epsilon$, where the dark fermion $\chi$ only constitutes a small fraction of the total DM relic abundance.

\end{itemize}

In summary, while standard benchmarks of GeV-scale fermionic DM coupled to the SM by a dark photon are experimentally excluded, scenarios with moderately small couplings $g_D \sim 10^{-1} - 10^{-2}$ are still allowed in the resonant annihilation regime and the remaining window of the parameter space can be probed in a complementary way by future direct detection and collider experiments.

\section*{Acknowledgements}

We thank Andrew Cheek for help with implementing this model in the RAPIDD code. We acknowledge support from the Spanish Agencia Estatal de Investigaci\'on through the grants PID2021-125331NB-I00, PID2021-124704NB-I00, and CEX2020-001007-S, funded by MCIN/AEI/10.13039/501100011033. DGC also acknowledges support from the Spanish Ministerio de Ciencia e Innovaci\'on (MCIN) under grant CNS2022-135702. JMN acknowledges support from the Spanish MCIN through the grant CNS2023-144536. PF acknowledges funding from the European Union’s Horizon Europe research and innovation programme under the Marie Skłodowska-Curie Staff Exchange grant agreement No 101086085 – ASYMMETRY. PF would like to thank the Massachusetts Institute of Technology (MIT) for hospitality during the completion of this work.
Feynman diagrams were drawn using {\sc TikZ-Feynman}~\cite{Ellis:2016jkw}.

\bibliographystyle{JHEP}
\bibliography{literature}

\end{document}